\documentclass[journal,draftclsnofoot,onecolumn,12pt]{IEEEtran}
\usepackage{amsthm,amssymb,graphicx,multirow,amsmath,color,amsfonts}%,ulem}
\usepackage[update,prepend]{epstopdf}
\usepackage[noadjust]{cite}
\usepackage[latin1]{inputenc}
\usepackage{tikz}
\usepackage{bbm} % for \mathbbm{1}
\usepackage{pdfpages}
\usepackage{tabulary}
\usepackage{multirow}
\usepackage{comment}
\usepackage{subfigure}
\usepackage{float}
\usepackage[titlenumbered,ruled]{algorithm2e}
\usepackage[noend]{algpseudocode}

\def\nb0{{\mathbf{0}}}
\def\nb1{{\mathbf{1}}}

\newtheorem{lemma}{Lemma}

\newtheorem{definition}{Definition}

\newtheorem{theorem}{Theorem}

\newtheorem{remark}{Remark}

\allowdisplaybreaks % Allows breaking of eqnarray over multiple pages (avoids unnecessary blanks in the document before eqnarray)
\begin{document}
\title{Coverage Analysis and Trajectory Optimization for Aerial Users with Dedicated Cellular Infrastructure}
%\title{Aerial Dedicated Infrastructure}
\author{
	Yujie Qin, Mustafa A. Kishk, {\em Member, IEEE}, and Mohamed-Slim Alouini, {\em Fellow, IEEE}
	\thanks{Yujie Qin and Mohamed-Slim Alouini are with Computer, Electrical and Mathematical Sciences and Engineering (CEMSE) Division, King Abdullah University of Science and Technology (KAUST), Thuwal, 23955-6900, Saudi Arabia. Mustafa Kishk is with the Department of Electronic Engineering, National University of Ireland, Maynooth, W23 F2H6, Ireland. (e-mail: yujie.qin@kaust.edu.sa; mustafa.kishk@mu.ie; slim.alouini@kaust.edu.sa).} % remove the date for conference drafts
	
}
%\date{\today}
\maketitle
\begin{abstract}
In this paper, we consider a novel cellular network for aerial users, which is composed of dedicated base stations (BSs), whose antennas are directed towards aerial users, and traditional terrestrial BSs (TBSs). Besides, the dedicated BSs are deployed on roadside furniture, such as lampposts and traffic lights, to achieve multiple features while occupying less space. Therefore, the locations of dedicated BSs and TBSs are modeled by a Poisson-line-Cox-process (PLCP) and Poisson point process (PPP), respectively. For the proposed network, we first compute the aerial coverage probability and show that the deployment of dedicated BSs improves the coverage probability in both high dense areas and rural areas. We then consider a cellular-connected UAV that has a flying mission and optimize its trajectory to maximize the minimal achievable signal-to-interference-plus-noise ratio (SINR) (Max-Min SINR). To obtain the Max-Min SINR and minimal time trajectory that satisfies the Max-Min SINR, we proposed two algorithms that are practical in large-scale networks. Finally, our results show that the optimal density of dedicated BSs which maximizes Max-Min SINR decreases with the increase of the road densities.
\end{abstract}
\section{Introduction}
Featured by high flexibility and mobility, unmanned aerial vehicles (UAVs), also known as drones, have drawn great attention in a broad range of civilian applications, such as transportation, package delivery, communications, and surveillance. For instance, UAVs are expected to help in delivery of goods and even passengers \cite{cvitanic2020drone}, security monitoring \cite{shakhatreh2019unmanned}, traffic control, and even finishing multiple tasks simultaneously \cite{khosravi2021multi,9933782}. Among all these appealing applications, it is of vital importance to ensure a reliable connection between UAVs and ground BSs, and only in this way, the safety of the public and residents can be guaranteed. Observing that the currently existing infrastructure is mainly designed for serving the ground users while the aerial users experience a sidelobe gain, which may be difficult to meet the goal of full coverage of future aerial networks. The idea of dedicated infrastructure for aerial users is expected to be a potential solution to improve the connectivity of aerial users \cite{9917390}.

However, building new infrastructure for aerial users is tricky and may cost a lot of money and energy. Generally, BSs are on-grid and connected by physical lines, to cooperate and share energy \cite{jung2021renewable}, and both the installation and maintenance costs are high. Besides, deploying a high density of new BSs in space-limited urban areas is inefficient.
Hence, we are seeking an alternative solution such as integrating the aerial dedicated antennas into existing infrastructure, such as roadside furniture.  Considering the lighting infrastructure and traffic signs as a commonly used source, deploying the dedicated antennas on such structures to form aerial dedicated BSs may be a suitable way to go:  instead of being connected by fiber, these dedicated BSs can be connected by wireless backhaul links and form a dense mesh network \cite{wirelessbackhaulref,wirelessbackhaulref1}. For instance, radio access nodes can be deployed on lampposts or rooftops and wirelessly connect with each other by using mmWave. Instead of connecting each node to the fiber point of presence (PoP), most of the nodes (distribution nodes) connect to the Fiber PoP indirectly via LoS links with other nodes. To ensure reliable connectivity and large capacity, high-dense deployment over large geographic areas is required, e.g., if some nodes are blocked due to some unpredictable reasons, data can be delivered through other LoS paths. Such a hybrid system provides multiple features with less space and without the need for new grid infrastructure, which saves a huge amount of money. Besides, such a system is not something unrealistic and it actually has been tested and implemented (for ground users only). For instance, Facebook/Meta terragraph technology \cite{nordrum2019facebook}, deploys highly-dense transmitters and receivers on the rooftops and light posts to achieve high-speed and low-cost transmission. Therefore, in this work, we seek the feasibility of integrating aerial dedicated BSs on roadside furniture for aerial users. However, due to the wireless connection feature, this network has lower capacity, higher latency, and higher cumulative traffic compared to the fiber backhaul.

Motivated by the concept of smart city, which aims to integrate information and communication technology into existing systems and achieves efficient networks, we study the performance of a novel cellular network which is composed of traditional TBSs and dedicated BSs that are deployed on the roadside furniture. To get system-level insights of the proposed network, we analyze the system performance from the perspective of static UAVs and UAVs with mobility, respectively. For static UAVs, we study the coverage probabilities, and for a moving UAV, e.g., a UAV that has a mission such as delivering a package, we optimize the trajectory and study the  Max-Min SINR during traveling. Specifically, we use tools from stochastic geometry to model the locations of TBSs and aerial dedicated BSs and optimization tools to design the UAVs' trajectories given the random locations.

\subsection{Related Work}
Literature related to this work can be categorized into: (i) incorporating aerial users into existing networks, (ii) optimization of UAV communication networks , and (iii)  stochastic geometry-based analysis of UAV networks. A brief discussion on related works is provided in the following lines.

{\em Incorporating aerial users into existing networks.}
An overview of UAV communication enabled by cellular networks was provided in \cite{8470897} and some discussions about network design and potential technologies were elaborated. Authors in \cite{7470932} discussed designing and implementing of future aerial networks and drew a conclusion that the technologies which are suitable for aerial networks are highly restricted by regulations and mechanical limitations.  Authors in \cite{8301389} investigated the performance  a typical rural aerial network and showed that when the load is high, aerial users experience a poor connection.  An altitude-dependent path loss exponent and fading function was analyzed in \cite{azari2017ultra}, in which the authors extended the models for G2G links to A2G channels. Authors in \cite{7842099} investigated the communication model for terrestrial-aerial communication channels. 

{\em Optimization-based UAV communication networks.} Authors in \cite{7888557}  proposed an energy-efficient trajectory design for UAVs given the constraints such as initial/final locations, velocities, and accelerations. In \cite{7762053}, authors optimized the placement of UAVs to ensure coverage for a group of ground users. A jointly optimization of UAV trajectory and transmit power was investigated in \cite{8068199} and results showed a lower outage probability. A high mobility UAV communication relay was analyzed in \cite{7572068}. The authors studied the throughput maximization problems by designing UAV trajectory and power allocations. Authors in \cite{8422584,8531711} investigated a Max-Min SNR based UAV trajectory optimization problem. Authors in \cite{8247211} jointly designed multiple UAVs' trajectories and transmit powers to maximize the throughput of a group of ground users. In \cite{8698468}, the authors considered a UAV that collects data from multiple wireless sensor networks and jointly optimized UAV's trajectory and communication scheduling to maximize the minimum average data collection rate. In \cite{8528463}, authors studied the reliability of command and control channel between UAVs and a traditional cellular network with massive MIMO. Besides, a deep reinforcement learning-based approach of optimizing multiple cellular-connected UAVs to minimize their interference with ground users was analyzed in \cite{8654727}. Authors in \cite{mei2019cellular} studied the uplink inter-cell interference coordination design for a cellular network simultaneously serving UAVs and ground users. Typically, the maximum weighted sum-rate of UAVs and users is jointly optimized based on power allocations and uplink cell associations. Authors in \cite{pang2019uplink} maximized the sum rate of UAV uplink transmission by optimizing the precoding vectors at the multi-antenna UAVs. A deep reinforcement learning-based intelligent navigation task of a cellular-connected UAV network was investigated in \cite{li2022path} which aims to minimize the weighted sum of time cost and expected outage duration alongside UAV trajectory.

{\em Stochastic geometry-based analysis of UAV networks.} Stochastic geometry is widely used to model large-scale communication networks \cite{7733098,6524460} and modeling the locations of UAVs as a Poisson point process (PPP) is commonly used in the literature \cite{zeng2016wireless,mozaffari2019tutorial,qin2020performance}. Authors in \cite{8833522,galkin2019stochastic,alzenad20173} studied the downlink coverage probability and averaged data rate by modeling the locations of TBSs and UAVs by two independent PPPs. Beside PPP, Matern cluster process (MCP) is another commonly used point process. Authors in \cite{9773146,7809177,qin2021influence} use this model to jointly model the locations of UAVs and users, where the UAVs are deployed above cluster centers and users are uniformly distributed in the cluster. Authors in \cite{9852974,8866716} studied the laser powered UAVs and studied the density of the laser beam directors to ensure energy coverage probability. Authors in \cite{9712177} studied a 3D two-hop cellular network with coexisting TBSs and UAVs to serve users. To guarantee line-of-sight (LoS) coverage to all users, authors in \cite{9419071} investigated a determined 3D placement and orientations problem of UAVs.

Unlike the existing literature which improves UAV networks' performance by optimizing the locations and trajectories, the proposed system improves the connectivity by deploying dedicated aerial BSs; and unlike the existing work which converts the current BSs into new BSs or simply deploys new BSs, the proposed system integrates the new BSs to existing infrastructure to save space and be energy efficient. Besides, we study the performance of moving UAVs based on stochastic geometry models, which is more practical due to the large-scale feature. Besides, we would like to clarify that the coverage probability of ground users decreases with the increasing density of aerial dedicated users due to the increasing interference, as discussed in our previous work \cite{9917390}. The differences between this work and our previous work are (i) in this work we consider a more complex system model in which dedicated BSs are modeled by a PLCP while in previous work the dedicated BSs are deployed by converting existing TBSs and (ii) since we optimize the UAV trajectory in this work, we simplify the antenna radiation pattern model, which can be considered as a special case in \cite{9917390}, to be feasible for optimization.

\subsection{Contribution}
This paper systematically investigates the feasibility and performance of integrating dedicated BSs into existing networks. Our main contributions of this work can be summarized below.

{\em Novel system model.}
To provide reliable connectivity for aerial users, we propose a novel wireless cellular network that is composed of TBSs and aerial dedicated BSs. To integrate this new infrastructure into existing networks, we consider these dedicated BSs to be deployed on the roadside furniture (install on the top and share the energy infrastructure, grid, battery, or solar panels), such as lampposts and traffic lights, to save space and be energy efficient. By doing so, aerial users, like UAVs, are able to experience mainlobe signals, hence, have better communication channels.

{\em Stochastic geometry-based UAV trajectory optimization.} Besides static UAVs, we also analyze the improvement of implementing the proposed network on the performance of a flying UAV. We consider a connectivity-constrained UAV which flies from the initial location to the destination and optimizes its trajectory to achieve the Max-Min SINR  during traveling. Besides, we propose an algorithm to obtain the minimal time trajectory that satisfies the Max-Min SINR constraint. Given the nature of stochastic geometry, which aims to analyze the performance of large-scale networks with a large number of BSs, this minimal time trajectory is a sub-optimal solution to the real case (the optimal solution is tricky to obtain due to the large number of BSs which results into infinite possible routes), however, it is more practical since the proposed system is more generic and focuses on the spatial statistics of the system rather than a particular system setup.

{\em System-level insights.} We use tools from stochastic geometry to model the locations of TBSs and dedicated BSs. Specifically, we model the locations of dedicated BSs by a Poisson-line-Cox-process (PLCP). Therefore, we are able to obtain the averaged system performance (averaged over the locations) and the proposed algorithms fit different scenarios. Our results show that the deployment of dedicated BSs improves the aerial coverage probability in both urban areas and rural areas, and the optimal value of point density varies with road densities. For moving UAVs, we show that the dedicated BSs improve the Max-Min SINR for different road densities (both rural and urban areas). We also observe an interesting trend in averaged minimal time of the trajectory: the averaged minimal time will increase first and then decrease with the increasing of the point densities. 

\section{System Model}
We consider a cellular-enabled UAV that can be connected to two types of ground BSs, traditional TBSs which mainly provide service to ground users, and dedicated BSs which are dedicated to aerial users \cite{9917390}, such as UAVs. Since the traditional TBSs have existed for decades while the new infrastructure, dedicated BSs, can be added to the existing street furniture \cite{eolpref},  the locations of the traditional TBSs are modeled by a PPP, $\Phi_{tb}$ with density $\lambda_{tb}$, and the locations of the dedicated BSs are modeled by a PLCP $\Phi_{db}$ with road density $\lambda_l$ and point density $\lambda_p$, since PLCP is commonly used in modeling the random nodes (dedicated BSs) on random lines (roads) \cite{dhillon2020poisson}, as shown in Fig. 1. Besides, to characterize the SINR, we let $\Phi_l$ denote the set of lines (roads) with density $\lambda_l$.  Firstly, for static UAVs hovering at fixed locations to provide a particular service, we are interested in the aerial coverage probability of this novel setup of PPP-modeled TBSs and PLCP-modeled dedicated aerial BSs.  Secondly, for traveling UAVs, we are interested in the maximum achievable minimum SINR (Max-Min SINR, $\gamma^{*}$) of the UAV trajectory and the minimum time of finishing the traveling mission. We assume that the reference UAV has a mission of flying from its initial location ${\rm S}$ to the final destination ${\rm D}$ which is $L$ km away and in a random direction. During the UAV's route, it flies at a fixed altitude $h_u$ and we want it to achieve a reliable connection with ground BSs, either TBSs or dedicated BSs. To ensure that, the minimum SINR of the whole trajectory should be above a predefined threshold: at any time of the UAV trajectory, the SINR should be greater than $\gamma^{*}$.
\begin{figure}[ht]
	\centering
	\includegraphics[width=0.7\columnwidth]{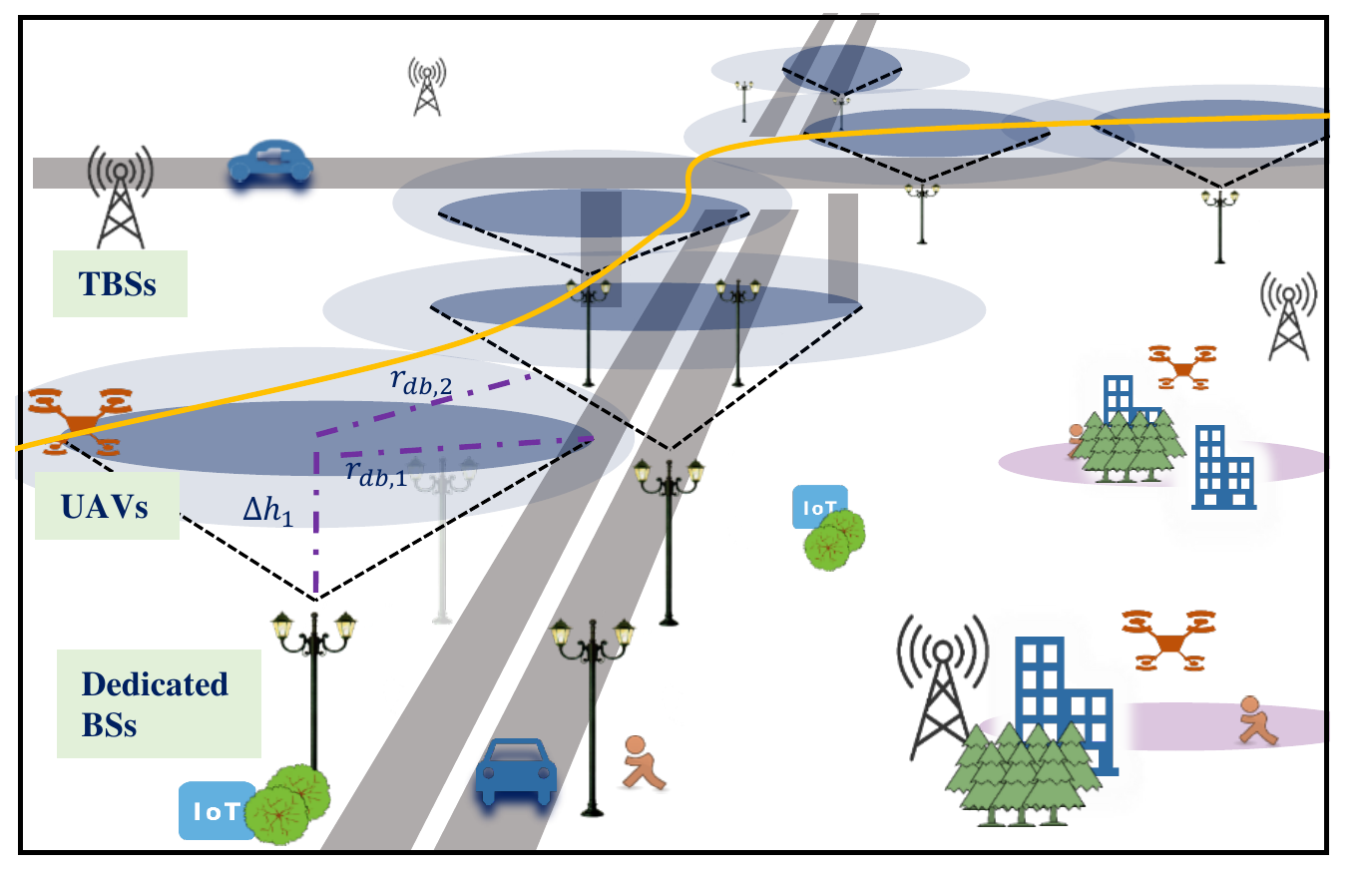}
	\caption{Illustration of the system model.}
	\label{fig_sysill}
\end{figure}

As shown in Fig. \ref{fig_sysill}, to ensure a reliable connectivity with aerial users, dedicated antennas are deployed on roadside furniture, such as lampposts. Instead of flying in a straight line between the initial point to the destination, the reference UAV adjusts its trajectory to achieve a target SINR. In the following part, we illustrate the proposed system from the perspective of aerial coverage probability and optimal trajectory, respectively.

\subsection{Communication Channel Model}
The detailed communication channel model of the dedicated BSs is defined in \cite{9917390}. To simplify the selection  of notation and have a feasible zone in the trajectory optimization part, the antenna gain when the aerial user associates with a dedicated BS in this work is 
\begin{align}
g_{a}(d) =  \left\{ 
\begin{aligned}
	g_{m} ,  & \quad d<z_{db}\\
	g_{s} ,  & \quad d>z_{db},\\
\end{aligned} \right.
\end{align}
where $d$ is the horizontal distance between the UAV and the dedicated BS and $z_{db}$ denotes the horizontal coverage range of the mainlobe of dedicated BSs, and $g_{m}$ and $g_{s}$ denote the mainlobe and sidelobe gain, respectively.

Let $\rho_{tb}$ and $\rho_{db}$ be the transmission power of TBSs and dedicated BSs, respectively. Given the horizontal distance between the UAV and the serving TBS/dedicated BS, $R_{tb}$ and $R_{db}$, respectively, the received power of the UAV is
\begin{align}
	\label{eq_pt_pdb}
p_{tb}(R_{tb}) = \left\{ 
\begin{aligned}
 p_{tb,l}(R_{tb,l}) = \eta_{l}\rho_{tb}g_{s}G_{l}D^{-\alpha_{l}}_{tb,l},  & \quad \text{in the case of LoS},\\
 p_{tb,n}(R_{tb,n}) =  \eta_{n}\rho_{tb}g_{s}G_{n}D^{-\alpha_{n}}_{tb,n},  & \quad \text{in the case of NLoS},\\
\end{aligned} \right.\nonumber\\
p_{db}(R_{db}) = \left\{ 
\begin{aligned}
	p_{db,l}(R_{db,l}) = g_{a}(R_{db,l})\eta_{l}\rho_{db}G_{l}D^{-\alpha_{l}}_{db,l},  & \quad \text{in the case of LoS},\\
	p_{db,n}(R_{db,n}) =  g_{a}(R_{db,n})\eta_{n}\rho_{db}G_{n}D^{-\alpha_{n}}_{db,n},  & \quad \text{in the case of NLoS},\\
\end{aligned} \right.
\end{align}
where $D_{tb,\{l,n\}}$ and $D_{db,\{l,n\}}$ are the Euclidean distances and obtained by $D_{tb,\{l,n\}} = \sqrt{\Delta h_{2}^2+R_{tb,\{l,n\}}^{2}}$ and $D_{db,\{l,n\}} = \sqrt{\Delta h_{1}^2+R_{db,\{l,n\}}^{2}}$, respectively, in which $R_{tb,l}$, $R_{tb,n}$, $R_{db,l}$ and $R_{db,n}$ denote the distance between the UAV and the serving LoS TBS, NLoS TBS, LoS dedicated BS and NLoS dedicated BS, respectively. $\Delta h_1 = h_{u}-h_{db}$ and $\Delta h_2 = h_{u}-h_{tb}$, where $h_{tb}$ and $h_{db}$ are the heights of the TBSs and dedicated BSs and $h_u$ is the altitude of UAVs, $\alpha_l$, $\eta_l$ and $G_l$ denote the path loss, additional loss and channel fading of LoS transmission and $\alpha_n$, $\eta_n$ and $G_n$ denote the path loss, additional loss and channel fading of NLoS transmission. 

Besides, the probability of establishing a LoS/NLoS link between the UAV and the serving BS is given in \cite{al2014optimal} as
\begin{align}
	P_l(r) &= \frac{1}{1+a\exp(-b(\frac{180}{\pi}\arctan(\frac{\Delta h}{r})-a))},\nonumber\\
	P_n(r) &= 1-P_l(r),\label{eq_LoSprob}
\end{align}
where $r\in\{R_{tb},R_{db}\}$, $\Delta h \in \{\Delta h_{1} = h_{u}-h_{tb},\Delta h_{2} = h_{u}-h_{db}\}$, and $a$ and $b$ are two environment variables.

Assuming that the reference UAV associates with the BS that provides the strongest average received power. We define the association probability with TBSs and dedicated BSs as follows.
\begin{definition}[Association Probability]
Let $\mathcal{A}_{db,l}(d,\theta_d)$, $\mathcal{A}_{db,n}(d,\theta_d)$, $\mathcal{A}_{tb,l}(d,\theta_d)$ and $\mathcal{A}_{tb,n}(d,\theta_d)$ be association probabilities between the reference UAV and the nearest LoS dedicated BS, NLoS dedicated BS, nearest LoS TBS, and NLoS TBS, respectively,
\begin{align}
	\mathcal{A}_{db,l}(d,\theta_d) &= \mathcal{A}_{db,l-db,n}(d,\theta_d)\mathcal{A}_{db,l-tb,l}(d)\mathcal{A}_{db,l-tb,n}(d),\nonumber\\
	\mathcal{A}_{db,n}(d,\theta_d) &= \mathcal{A}_{db,n-db,l}(d,\theta_d)\mathcal{A}_{db,n-tb,l}(d)\mathcal{A}_{db,n-tb,n}(d),\nonumber\\
	\mathcal{A}_{tb,l}(d,\theta_d) &= \mathcal{A}_{tb,l-db,n}(d,\theta_d)\mathcal{A}_{tb,l-db,l}(d,\theta_d)\mathcal{A}_{tb,l-tb,n}(d),\nonumber\\
	\mathcal{A}_{tb,n}(d,\theta_d) &= \mathcal{A}_{tb,n-db,l}(d,\theta_d)\mathcal{A}_{tb,n-tb,l}(d)\mathcal{A}_{tb,n-db,n}(d,\theta_d),
\end{align}
where $d$ denotes the horizontal communication distance and $\theta_d = \arcsin(\frac{y_l}{d})$, in which $y_l$ is the distance to the line from the PLCP that contains the serving or interfering dedicated BS (more details are provided in Lemma \ref{lemma_distancedistribution} and Lemma \ref{lemma_associateprob} and Fig. \ref{fig_ill_I}), and $\mathcal{A}_{a-b}(d,\theta_d)$ denotes the probability that the average received power from BS $a$ is stronger than the average received power from BS $b$.
\end{definition}

As mentioned, we are interested in the aerial coverage probability of the proposed network. The coverage probability is defined as the probability that a typical user is successfully served by the BS, which refers to the probability that its SINR is greater than some predefined threshold.
\begin{definition}[Coverage Probability]
Let $\tau$ be the SINR threshold, the coverage probability is defined as 
\begin{align}
P_{\rm cov} = P_{\rm cov,tb}+P_{\rm cov,db},
\end{align}
where $P_{\rm cov,tb}$ and $P_{\rm cov,db}$ denote the coverage probability when the reference UAV associates with TBS and dedicated BS, respectively,
\begin{align}
	\label{eq_def_cov}
P_{\rm cov,tb} &= \mathbb{E}_{R_{tb,l},\theta_d}[\mathcal{A}_{tb,l}(R_{tb,l},\theta_d)\mathbb{P}({\rm SINR_{tb,l}}(R_{tb,l})>\tau)]+\mathbb{E}_{R_{tb,n},\theta_d}[\mathcal{A}_{tb,n}(R_{tb,n},\theta_d)\mathbb{P}({\rm SINR_{tb,n}}(R_{tb,n})>\tau)],\nonumber\\
P_{\rm cov,db} &= \mathbb{E}_{R_{db,l},\theta_d}[\mathcal{A}_{db,l}(R_{db,l},\theta_d)\mathbb{P}({\rm SINR_{db,l}}(R_{db,l})>\tau)]\nonumber\\
&\quad+\mathbb{E}_{R_{db,n},\theta_d}[\mathcal{A}_{db,n}(R_{db,n},\theta_d)\mathbb{P}({\rm SINR_{db,n}}(R_{db,n})>\tau)],
\end{align}
in which SINR  and the aggregated interference are, respectively, given by
\begin{align}
{\rm SINR}_{\{tb,db\}}(\{R_{tb},R_{db}\}) &= \frac{p_{\{tb,db\}}(R_{\{tb,db\}})}{\sigma^{2}+I_{\{tb,db\}}},\label{eq_SINR}\\
I_{\{tb,db\}} &= \sum_{tb_l,i\in \Phi_{tb,l}\setminus t_0}\eta_{l}\rho_{tb}g_{s}G_{l,tb_l,i}D_{tb_l,i}^{-\alpha_{l}}+\sum_{tb_n,j\in \Phi_{tb,n}\setminus t_0}\eta_{n}\rho_{tb}g_{s}G_{n,tb_n,k}D_{tb_n,k}^{-\alpha_{n}}\nonumber\\
&+\sum_{db_l,k\in \Phi_{db}\setminus t_0}\eta_{l}g_a(\sqrt{D_{db_l,k}^2-\Delta h_{1}^{2}})\rho_{db}G_{l,db_l,k}D_{db_l,k}^{-\alpha_{l}}\nonumber\\
&+\sum_{db_n,m\in \Phi_{db,n}\setminus t_0}\eta_{n}g_a(\sqrt{D_{db_n,m}^2-\Delta h_{1}^{2}})\rho_{db}G_{n,db_n,m}D_{db_n,m}^{-\alpha_{n}},
\end{align}
where $t_0$ denotes the location of the serving BS, $\Phi_{tb,l}$ and $\Phi_{tb,n}$ are subsets of $\Phi_{tb}$ denote the locations of the LoS TBS and NLoS TBSs,  $\Phi_{db,l}$ and $\Phi_{db,n}$ are subsets of $\Phi_{db}$ denote the locations of the LoS and NLoS dedicated BSs, respectively, and the subscript $\{l,n\}$ denotes the LoS or NLoS transmission, as defined in (\ref{eq_pt_pdb}).
\end{definition}

\subsection{Trajectory Design}
As mentioned, we are also interested in the Max-Min SINR $\gamma^*$ that the reference UAV can achieve when it flies from a starting point to a destination and the minimal time given the target $\gamma^*$. Let $\mathbf{u}(t)$ be the trajectory of UAV with S and D denoting the locations of the initial location and the destination of the UAV. Let $T$ be defined as the time of finishing the traveling mission from ${\rm S}$ to ${\rm D}$ and $\gamma^t$ denotes the mean SINR achieved by UAV at time $t$, in which $0< t\leq T$.

We first observe that the mean SINR $\gamma^t$ at time $t$ is a function of the serving distance (distance to the serving BS). Conditioned on the serving BS being a dedicated BS or TBS, the mean SINR given the horizontal communication distance $d$ are, respectively, given by
	\begin{align}
		\label{eq_def_SINR}
		{\gamma}_{db}(d) &=\mathbb{E}_{\theta_d} \bigg[\int_{0}^{\infty}P_{l}(d)P_{s,db,l}(\tau,d,\theta_d)+P_{n}(d)P_{s,db,n}(\tau,d,\theta_d){\rm d}\tau\bigg],\nonumber\\
		{\gamma}_{tb}(d) &=\mathbb{E}_{\theta_d} \bigg[\int_{0}^{\infty} P_{l}(d) P_{s,tb,l}(\tau,d,\theta_d) +P_{n}(d) P_{s,tb,n}(\tau,d,\theta_d){\rm d}\tau \bigg],
	\end{align}
	where $P_{s,\{\cdot\}}(\tau,d,\theta_d)$ is defined as the conditional success probability and obtained by
	\begin{align}
		P_{s,\{\cdot\}}(\tau,d,\theta_d) = \mathbb{P}({\rm SINR_{\{\cdot\}}}(d)>\tau),
	\end{align}
in which ${\rm SINR_{\{\cdot\}}}(d)$ is given in (\ref{eq_SINR}) and the subscript $\{l,n\}$ in $P_{s,\{\cdot\}}(\tau,d)$ denotes the LoS or NLoS transmission, as defined in (\ref{eq_pt_pdb}).

\begin{definition}[Max-Min SINR]
Let $\bar{\gamma}$ be the minimum SINR during a UAV trajectory and $\gamma^*$ is the Max-Min SINR of the UAV trajectory given an arbitrary realization of locations of TBSs and dedicated BSs,
\begin{align}
\mathcal{P}_{1}:\quad & \gamma^{*} = \max_{\mathbf{u}(t)} \bar{\gamma},\nonumber\\
{\rm s.t.}\quad &\gamma^t = \max(\gamma_{tb}(d),\gamma_{db}(d)),\nonumber\\
& \gamma^t \geq \bar{\gamma},\quad \forall\quad 0< t\leq T,
\end{align}
where $\gamma^t$ denotes the mean SINR achieved by UAV at time $t$.
\end{definition}

 Let $r_{tb}({\gamma})$ and $r_{db}(\gamma)$ be the maximal communication distances between the reference UAV and the serving TBS and dedicated BS that achieve a target $\gamma$, respectively, where $\gamma$ denotes a value of mean SINR. The $r_{tb}({\gamma})$ and $r_{db}(\gamma)$ can be obtained by solving the inverse function of (\ref{eq_def_SINR}).
 
 Consequently, the optimization problem of the minimal time trajectory given $\gamma^*$ is formulated as
\begin{align}
	\mathcal{P}_2 : &\min_{\mathbf{u}(t)} T \nonumber\\
	&{\rm s.t.}\quad \mathbf{u}(0) = {\rm S},\quad \mathbf{u}(T) = {\rm D},\\
	&\qquad \mathbf{R}_{tb} \leq r_{tb}({\gamma}^{*}), \quad \mathbf{R}_{db} \leq r_{db}(\gamma^{*}),\label{eq_P1_SINR_constraint}
\end{align}
where $\mathbf{R}_{tb}$ and $\mathbf{R}_{db}$ denote the distances between the reference UAV and the serving TBSs and dedicated BSs along its trajectory, respectively. Besides, we would like to mention that the initial location and the destination of UAVs are totally random, which means that the UAVs are not necessarily to start exactly above a BS and ends exactly above a BS.

\section{Coverage Probability}
In this part, we provide the analysis for the aerial coverage probability. Before deriving the equations for the coverage probability, we first need to compute the association probability and some distance distributions, such as the PDFs of $R_{tb}$ and $R_{db}$.

\subsection{Distance Distribution}
Recall that $R_{tb,l}$, $R_{tb,n}$, $R_{db,l}$ and $R_{db,n}$ are the distances to the nearest LoS TBS, NLoS TBS, LoS dedicated BS and NLoS dedicated BS, respectively. The probability density functions (PDFs) and cumulative density functions (CDFs) of the above mentioned distances are provided in the following lemma.
\begin{lemma}[PDFs and CDFs of the $R_{tb,l}$, $R_{tb,n}$, $R_{db,l}$ and $R_{db,n}$]
	\label{lemma_distancedistribution}
	The PDFs of the $R_{tb,l}$, $R_{tb,n}$, $R_{db,l}$ and $R_{db,n}$ are, respectively, given by
\begin{align}
	\label{eq_pdf}
	f_{R_{db,\{l,n\}}}(r) = &\bigg[2\pi P_{\{l,n\}}(r)\cos\theta_d\nonumber\\
	&+ 2\pi\lambda_l\int_{0}^{r}2\lambda_p\frac{ r}{\sqrt{r^2-\rho^2}}P_{\{l,n\}}(r)\exp\bigg(-2\lambda_p\int_{0}^{\sqrt{r^2-\rho^2}}P_{\{l,n\}}(\sqrt{\rho^2+z^2}){\rm d}z\bigg){\rm d}\rho\bigg]\nonumber\\
	&\exp\bigg(-2\lambda_p\int_{0}^{r\cos(\theta_d)}P_{\{l,n\}}(\sqrt{z^2+r^2\sin^{2}(\theta_d)}){\rm d}z\nonumber\\
	&\qquad-2\pi\lambda_l\int_{0}^{r}1-\exp\bigg(-2\lambda_p\int_{0}^{\sqrt{r^2-\rho^2}}P_{\{l,n\}}(\sqrt{\rho^2+z^2}){\rm d}z\bigg){\rm d}\rho\bigg),\nonumber\\
	f_{R_{tb,\{l,n\}}}(r) =&  2\pi\lambda_{tb} P_{\{l,n\}}(r)\exp\bigg(-2\pi\lambda_{tb}\int_{0}^{r}zP_{\{l,n\}}(z){\rm d}z\bigg),
\end{align}
consequently, the corresponding CDFs are obtained by taking the integration and given by
\begin{align}
	\label{eq_CDF}
	F_{R_{db,\{l,n\}}}(r,\theta_d) &= 1-\exp\bigg(-2\lambda_p\int_{0}^{r\cos(\theta_d)}P_{\{l,n\}}(\sqrt{z^2+r^2\sin^{2}(\theta_d)}){\rm d}z\nonumber\\
	&\qquad-2\pi\lambda_l\int_{0}^{r}1-\exp\bigg(-2\lambda_p\int_{0}^{\sqrt{r^2-\rho^2}}P_{\{l,n\}}(\sqrt{\rho^2+z^2}){\rm d}z\bigg){\rm d}\rho\bigg),\nonumber\\
	F_{R_{tb,\{l,n\}}}(r) &= 1-\exp\bigg(-2\pi\lambda_{tb}\int_{0}^{r}zP_{\{l,n\}}(z){\rm d}z\bigg),
\end{align}
in which $\theta_d = \arcsin(\frac{y_l}{r})$ as shown in Fig. \ref{fig_I_int} and we need this variable to compute the lower bound of integration in the Laplace transform of interference in (\ref{eq_Intlowerbound}).
\end{lemma}

Recall that we assume that the reference UAV associates with the BS that provides the strongest average received power and $\mathcal{A}_{db,l}(d,\theta_d)$, $\mathcal{A}_{db,n}(d,\theta_d)$, $\mathcal{A}_{tb,l}(d,\theta_d)$ and $\mathcal{A}_{tb,n}(d,\theta_d)$ are association probabilities that the reference UAV associates with the nearest LoS dedicated BS, NLoS dedicated BS, nearest LoS TBS and NLoS TBS at distance $d$ away, respectively.

\begin{lemma}[Association Probability]
	\label{lemma_associateprob}
	The association probabilities given the horizontal communication distance, are given by
	\begin{align}
		\label{eq_associateprob}
		\mathcal{A}_{db,l}(d,\theta_d) &= \mathcal{A}_{db,l-db,n}(d,\theta_d)\mathcal{A}_{db,l-tb,l}(d)\mathcal{A}_{db,l-tb,n}(d)\nonumber\\
		&= \bar{F}_{R_{db,n}}(d_{db,db,ln}(d),\theta_d)\bar{F}_{R_{tb,l}}(d_{db,tb,ll}(d))\bar{F}_{R_{tb,n}}(d_{db,tb,ln}(d)),\nonumber\\
		\mathcal{A}_{db,n}(d,\theta_d) &= \mathcal{A}_{db,n-db,l}(d,\theta_d)\mathcal{A}_{db,n-tb,l}(d)\mathcal{A}_{db,n-tb,n}(d)\nonumber\\
		&= \bar{F}_{R_{db,l}}(d_{db,db,nl}(d),\theta_d)\bar{F}_{R_{tb,l}}(d_{db,tb,nl}(d))\bar{F}_{R_{tb,n}}(d_{db,tb,nn}(d)),\nonumber\\
		\mathcal{A}_{tb,l}(d,\theta_d) &= \mathcal{A}_{tb,l-db,n}(d)\mathcal{A}_{tb,l-db,l}(d,\theta_d)\mathcal{A}_{tb,l-tb,n}(d)\nonumber\\
		&= \bar{F}_{R_{db,n}}(d_{tb,db,ln}(d),\theta_d)\bar{F}_{R_{db,l}}(d_{tb,db,ll}(d),\theta_d)\bar{F}_{R_{tb,n}}(d_{tb,tb,ln}(d)),\nonumber\\
		\mathcal{A}_{tb,n}(d,\theta_d) &= \mathcal{A}_{tb,n-db,l}(d,\theta_d)\mathcal{A}_{tb,n-tb,l}(d)\mathcal{A}_{tb,n-db,n}(d,\theta_d)\nonumber\\
		&= \bar{F}_{R_{db,l}}(d_{tb,db,nl}(d),\theta_d)\bar{F}_{R_{db,n}}(d_{tb,db,nn}(d),\theta_d)\bar{F}_{R_{tb,l}}(d_{tb,tb,nl}(d)),
	\end{align}
where $\bar{F}_{\{\cdot\}}(\cdot)$ denotes the complementary cumulative density function (CCDF) which equals to $1-F_{\{\cdot\}}(\cdot)$, where $F_{\{\cdot\}}(\cdot)$ is the CDF defined in (\ref{eq_CDF}) in Lemma \ref{lemma_distancedistribution} and the horizontal distances $d_{\{\cdot\}}$ are provided in Appendix \ref{app_d}.
\end{lemma}

\subsection{Laplace Transform}
Now we derive the Laplace transform for the aggregate interference in Lemma 3 later in this subsection.
Recall that we model the locations of dedicated BSs as a PLCP, which results in a slight difference in deriving the Laplace transform of the aggregate interference compared to the interference of a PPP. 

\begin{figure}[ht]
	\centering
	\includegraphics[width=0.9\columnwidth]{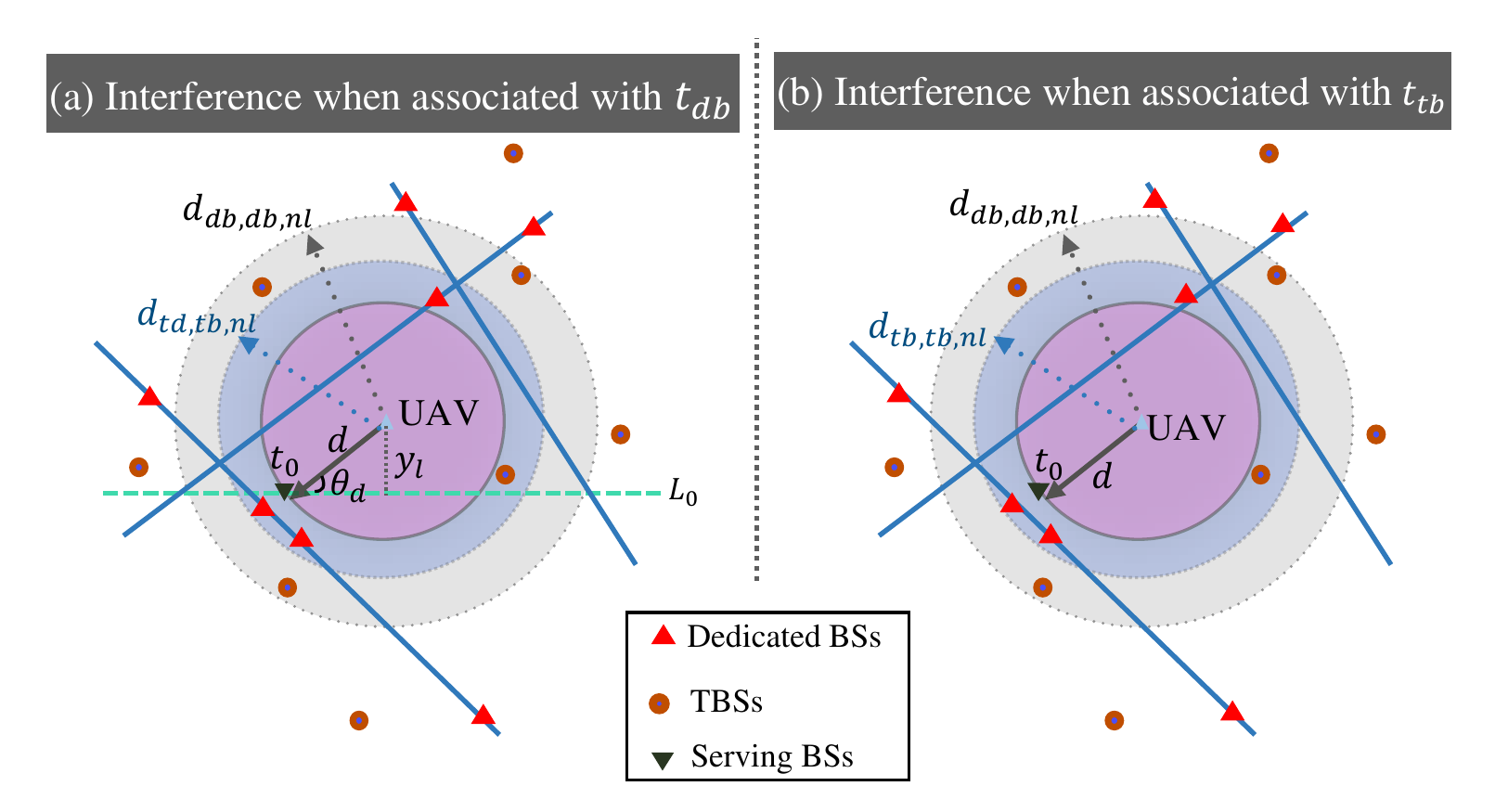}
	\caption{Illustration of interfers when the reference UAV associates with \textbf{(a)} a dedicated BS, \textbf{(b)} a TBS.}
	\label{fig_ill_I}
\end{figure}

When the reference UAV associates with a dedicated BS, a typical line from the PLP which passes through the serving dedicated BS exists. While when the reference UAV associates with a TBS, this typical line does not exist, as shown in Fig. \ref{fig_ill_I}. Hence, we discuss these two scenarios separately.

As shown in Fig. \ref{fig_ill_I} (a), when the reference UAV associates with a dedicated BS, the aggregate interference is composed of three sets, (i) the interference $I_{db,1}$ from the typical line $L_0$, (ii)
the interference $I_{db,2}$  from the rest of the lines $\Phi_{l}^{'}$, in which $\Phi_{l}^{'} =\Phi_l\backslash L_0$ denotes the set of lines, and (iii) the interference from TBSs. Therefore, the aggregate interference $I_{tb}$  when associating with a dedicated BS is 
\begin{align}
I_{db} &= I_{db,1}+I_{db,2}+I_{tb}\nonumber\\
&= I_{db,1,l}+I_{db,1,n}+I_{db,2,l}+I_{db,2,n}+I_{tb,l}+I_{tb,n}\nonumber\\
&\stackrel{(a)}{=} \sum_{db_{l},i\in L_0\setminus t_0}p_{db} g_a(Z_{db_{l},i}) \eta_l G_{l,db_{l},i} (Z_{db_{l},i}^2+\Delta h_{1}^2)^{-\alpha_l/2}\nonumber\\
&+\sum_{db_{n},j\in L_0\setminus t_0}p_{db} g_a(Z_{db_{n},j}) \eta_n G_{n,db_{n},j} (Z_{db_{n},j}^2+\Delta h_{1}^2)^{-\alpha_n/2}\nonumber\\
&+ \sum_{db_{l},i\in \Phi_{l}^{'}}p_{db} g_a(Z_{db_{l},i}) \eta_l G_{l,db_{l},i} (Z_{db_{l},i}^2+\Delta h_{1}^2)^{-\alpha_l/2}+\sum_{db_{n},j\in \Phi_{l}^{'}}p_{db} g_a(Z_{db_{n},j}) \eta_n G_{n,db_{n},j} (Z_{db_{n},j}^2+\Delta h_{1}^2)^{-\alpha_n/2}\nonumber\\
&+ \sum_{tb_{l,i}\in \Phi_{tb,l}}p_{tb}\eta_l g_{s}G_{l,tb_{l},i} D_{tb_{l},i}^{-\alpha_l}+\sum_{tb_{n},j\in \Phi_{tb,n}}p_{tb}\eta_n g_{s}G_{n,tb_{n},j} D_{tb_{n},j}^{-\alpha_n},
\end{align}
where  step (a) follows from the fact that the line process can be represented by $\Phi_{l} = \Phi_{l}^{'}\cup L_0$, $Z_{db,\{\cdot\}}$ denotes the horizontal distance between the reference UAV and the interfering dedicated BSs, and $t_0$ denotes the location of the serving BS.

As shown in Fig. \ref{fig_ill_I} (b), when the reference UAV associates with a TBS, the aggregate interference is given by
\begin{align}
	I_{tb} &= I_{db,2}+I_{tb}\nonumber\\
	&= I_{db,2,l}+I_{db,2,n}+I_{tb,l}+I_{tb,n}\nonumber\\
	&=\sum_{db_{l},i\in \Phi_{l}}p_{db} g_a(Z_{db_{l},i}) \eta_l G_{l,db_{l},i} (Z_{db_{l},i}^2+\Delta h_{1}^2)^{-\alpha_l/2}+\sum_{db_{n},j\in \Phi_{l}}p_{db} g_a(Z_{db_{n},j}) \eta_n G_{n,db_{n},j} (Z_{db_{n},j}^2+\Delta h_{1}^2)^{-\alpha_n/2}\nonumber\\
	&+ \sum_{tb_{l},i\in \Phi_{tb,l}\setminus t_0}p_{tb}\eta_l g_{s}G_{l,tb_{l},i} D_{tb_{l},i}^{-\alpha_l}+\sum_{tb_{n},j\in \Phi_{tb,n}\setminus t_0}p_{tb}\eta_n g_{s}G_{n,tb_{n},j} D_{tb_{n},j}^{-\alpha_n},
\end{align}
where $t_0$ denotes the location of the serving BS.

\begin{lemma}[Laplace Transform of The Interference]
	\label{lemma_laplace}
The Laplace transform the the aggregate interference is given by 
\begin{align}
\mathcal{L}_{I_{db}}(s,\theta_d) &= \mathcal{L}_{\rm gen}(s)\mathcal{L}_{\rm typ}(s,\theta_d),\nonumber\\
\mathcal{L}_{I_{tb}}(s) &= \mathcal{L}_{\rm gen}(s), 
\end{align}
where $\mathcal{L}_{\rm gen}(s)$ and $\mathcal{L}_{\rm typ}(s,\theta_d)$ denote the Laplace transform of the interference from general BSs, including TBSs and dedicated BSs on general lines, and typical dedicated BSs (dedicated BSs on $L_0$), respectively,
\begin{align}
&\mathcal{L}_{\rm gen}(s) = \exp\bigg(-2\pi\lambda_{tb} \sum_{c = \{l,n\}}\int_{v_{tb,c}(d)}^{\infty}(1-\kappa_1(c,s,z)) zP_c(z){\rm d}z \bigg)\nonumber\\
&\times\prod_{c = \{l,n\}} \exp\bigg(-2\pi\lambda_l \int_{0}^{v_{1,c}(d)}1-\exp\bigg(-2\lambda_p \int_{\sqrt{v_{1,c}^2(d)-\rho^2}}^{\infty}(1-\kappa_2(c,s,z))P_c(\sqrt{u^2+\rho^2}){\rm d}u\bigg){\rm d}\rho\bigg)\nonumber\\
& \times\prod_{c = \{l,n\}}\exp\bigg(-2\pi\lambda_l \int_{v_{1,c}(d)}^{\infty}1-\exp\bigg(-2\lambda_p \int_{0}^{\infty}(1-\kappa_2(c,s,z))P_c(\sqrt{u^2+\rho^2}){\rm d}u\bigg){\rm d}\rho\bigg),\\
&\mathcal{L}_{\rm typ}(s,\theta_d)= 
	\exp\bigg(-2\lambda_p\sum_{c = \{l,n\}} \int_{v_{0,c}(d)}^{\infty}(1-\kappa_3(c,s,z))P_c(x){\rm d}x \bigg),
\end{align}
where
\begin{align}
\kappa_1(c,s,z) &= \bigg(\frac{m_c}{m_c+s p_{tb}g_{s}\eta_c (z^2+\Delta h_{2}^2)^{-\alpha_{c}/2}}\bigg)^{m_c},\nonumber\\
\kappa_2(c,s,z) &= \bigg(\frac{m_c}{m_c+s p_{db} g_a(\sqrt{u^2+\rho^2}) \eta_c (u^2+\rho^2+\Delta h_{1}^2)^{-\alpha_c/2}}\bigg)^{m_c}\nonumber\\
\kappa_3(c,s,z) &= \bigg[\frac{m_c}{m_c+s p_{db} g_a(x) \eta_c (x^2+\Delta h_{1}^2)^{-\alpha_c/2}}\bigg]^{m_c},
\end{align}
and, 
\begin{align}
	\label{eq_Intlowerbound}
v_{0,l}(d) &=\left\{ 
\begin{aligned}
	d\cos\theta_d,&\quad t_0\in\Phi_{db,l},\\
	\sqrt{d_{db,db,nl}^2(d)-(d\sin\theta_d)^2}, & \quad t_0\in\Phi_{db,n},\\
\end{aligned} \right.\nonumber\\
v_{0,n}(d) &=\left\{ 
\begin{aligned}
\sqrt{(d_{db,db,ln}^2(d)-(d\sin\theta_d)^2)^{+}},&\quad t_0\in\Phi_{db,l},\\
d\cos\theta_d & \quad t_0\in\Phi_{db,n},\\
\end{aligned} \right.	\nonumber\\
v_{tb,\{l,n\}}(d) &=\left\{ 
\begin{aligned}
	d_{db,tb,l\{l,n\}}(d),&\quad t_0\in\Phi_{db,l},\\
	d_{db,tb,n\{l,n\}}(d), & \quad t_0\in\Phi_{db,n},\\
	d_{tb,tb,l\{l,n\}}(d), & \quad t_0\in\Phi_{tb,l},\\
	d_{tb,tb,n\{l,n\}}(d), & \quad t_0\in\Phi_{tb,n},\\
\end{aligned} \right.
%v_{tb,n}(d) =\left\{ 
%\begin{aligned}
%	d_{db,tb,ln}(d) & \quad t_0\in\Phi_{db,l},\\
%	d_{db,tb,nn}(d) & \quad t_0\in\Phi_{db,n},\\
%	d_{tb,tb,ln}(d), & \quad t_0\in\Phi_{tb,l},\\
%	d_{tb,tb,nn}(d), & \quad t_0\in\Phi_{tb,n},\\
%\end{aligned} \right.			\nonumber\\
v_{1,\{l,n\}}(d)  =\left\{ 
\begin{aligned}
d_{db,db,l\{l,n\}}(d),&\quad t_0\in\Phi_{db,l},\\
d_{db,db,n\{l,n\}}(d), & \quad t_0\in\Phi_{db,n},\\
d_{tb,db,l\{l,n\}}(d), & \quad t_0\in\Phi_{tb,l},\\
d_{tb,db,n\{l,n\}}(d), & \quad t_0\in\Phi_{tb,n},\\
\end{aligned} \right.
%v_{1,n}(d) =\left\{ 
%\begin{aligned}
%d_{db,db,ln}(d), & \quad t_0\in\Phi_{db,l},\\
%d_{db,db,nn}(d), & \quad t_0\in\Phi_{db,n},\\
%d_{tb,db,ln}(d), & \quad t_0\in\Phi_{tb,l},\\
%d_{tb,db,nn}(d), & \quad t_0\in\Phi_{tb,n},\\
%\end{aligned} \right.	
\end{align}
where  the distances $d_{\{\cdot\}}$ are defined in Lemma \ref{lemma_associateprob}.
\end{lemma}
\begin{IEEEproof}
See Appendix \ref{app_laplace}.
\end{IEEEproof}

Now we are able to obtain the coverage probability of this network. Observe from (\ref{eq_def_cov}) that coverage probability can be represented using the success probability,
\begin{align}
	P_{\rm cov,db} &= \mathbb{E}_{R_{db,l}}[\mathcal{A}_{db,l}(R_{db,l},\theta_d)\mathbb{P}({\rm SINR_{db,l}}(R_{db,l})>\tau)]+\mathbb{E}_{R_{db,n}}[\mathcal{A}_{db,n}(R_{db,n},\theta_d)\mathbb{P}({\rm SINR_{db,n}}(R_{db,n})>\tau)],\nonumber\\
	&=\mathbb{E}_{R_{db,l}}[\mathcal{A}_{db,l}(R_{db,l},\theta_d)P_{s,db_l}(\tau,d,\theta_d)]+
	\mathbb{E}_{R_{db,n}}[\mathcal{A}_{db,n}(R_{db,n},\theta_d)P_{s,db_n}(\tau,d,\theta_d)],\\
	P_{\rm cov,tb} &= \mathbb{E}_{R_{tb,l}}[\mathcal{A}_{tb,l}(R_{tb,l},\theta_d)\mathbb{P}({\rm SINR_{tb,l}}(R_{tb,l})>\tau)]+\mathbb{E}_{R_{tb,n}}[\mathcal{A}_{tb,n}(R_{tb,n})\mathbb{P}({\rm SINR_{tb,n}}(R_{tb,n})>\tau)]\nonumber\\
	 &= \mathbb{E}_{R_{tb,l}}[\mathcal{A}_{tb,l}(R_{tb,l},\theta_d)P_{s,t_l}(\tau,d)]
	 +\mathbb{E}_{R_{tb,n}}[\mathcal{A}_{tb,n}(R_{tb,n},\theta_d)P_{s,t_n}(\tau,d)],
\end{align}
 and we need the success probability to compute the mean SINR, as defined in (\ref{eq_def_SINR}). Hence, we first derive the success probability in the following lemma.

\begin{lemma}[Conditional Success Probability]
	\label{lemma_successprob}
 The conditional success probabilities given the reference UAV associates with a LoS dedicated BS, NLoS dedicated BS, LoS TBS and NLoS TBS, respectively, are given by
	\begin{align}
	\label{eq_successprob}
P_{s,db,{\{l,n\}}}(\tau,d,\theta_d) &= \sum_{k = 1}^{m_{\{l,n\}}}\binom{m_{\{l,n\}}}{k}(-1)^{k+1}\mathcal{L}_{I_{db,\{l,n\}}+\sigma^2}(\xi_{db,\{l,n\}}(\tau,d),\theta_d),\nonumber\\
%P_{s,db_n}(\tau,d,\theta_d) &= \sum_{k = 1}^{m_n}\binom{k}{m_n}(-1)^{k+1}\mathcal{L}_{I_{db,n}+\sigma^2}(k \beta_2(m_n)m_n \tau (p_{tb} \eta_n g_{a}(d))^{-1} (d^2+\Delta h_{1}^2)^{\frac{\alpha_n}{2}},\theta_d),\nonumber\\
P_{s,tb,{\{l,n\}}}(\tau,d) &= \sum_{k = 1}^{m_{\{l,n\}}}\binom{m_{\{l,n\}}}{k}(-1)^{k+1}\mathcal{L}_{I_{tb, \{l,n\} }+\sigma^2}(\xi_{tb,\{l,n\}}(\tau,d)), 
%P_{s,t_n}(\tau,d) &= \sum_{k = 1}^{m_n}\binom{k}{m_n}(-1)^{k+1}\mathcal{L}_{I_{tb,n}+\sigma^2}(k \beta_2(m_n)m_n \tau (p_{tb}g_{s} \eta_n )^{-1} (d^2+\Delta h_{2}^2)^{\frac{\alpha_n}{2}}),
\end{align}
where 
\begin{align}
	\xi_{db,\{l,n\}}(\tau,d) &= k \beta_2(m_{\{l,n\}})m_{\{l,n\}} \tau (p_{db} \eta_{\{l,n\}} g_{a}(d))^{-1} (d^2+\Delta h_{1}^2)^{\frac{\alpha_{\{l,n\}}}{2}},\nonumber\\
	\xi_{tb,\{l,n\}}(\tau,d) &= k \beta_2(m_{\{l,n\}})m_{\{l,n\}} \tau (p_{tb}g_{s} \eta_{\{l,n\}})^{-1} (d^2+\Delta h_{2}^2)^{\frac{\alpha_{\{l,n\}}}{2}}.
\end{align}
in which the subscript $l,n$ in the Laplace transform of the interference term denotes the LoS or NLoS transmission link between the UAV and serving BS, which influences the lower bound of the integration of the interference, as defined in (\ref{eq_Intlowerbound}), and $\beta_2(m) = (m!)^{-1/m}$ when $m \geq 1$ and this term appears due to the use of the upper bound approximation of Gamma distribution, which can be found in \cite{alzer1997some,8833522}.
\end{lemma}

Now we obtain the final expression of the coverage probability of the proposed network.
\begin{theorem}[Coverage Probability]
The coverage probability is given by
\begin{align}
P_{\rm cov} &= P_{\rm cov,tb}+P_{\rm cov,db}\nonumber\\
&= P_{\rm cov,tb,l}+P_{\rm cov,tb,n}+P_{\rm cov,db,l}+P_{\rm cov,db,n},
\end{align}
where,
\begin{align}
P_{\rm cov,db,\{l,n\}} &=\mathbb{E}_{\theta_d}\bigg[ \int_{0}^{\infty}\mathcal{A}_{db,\{l,n\}}(d,\theta_d)P_{s,db,\{l,n\}}(\tau,d,\theta_d)f_{R_{db,\{l,n\}}}(r,\theta_d){\rm d}r\bigg],\nonumber\\
%P_{\rm cov,db_n} &=\mathbb{E}_{\theta_d}\bigg[ \int_{0}^{\infty}\mathcal{A}_{db,n}(d,\theta_d)P_{s,db,n}(\tau,d,\theta_d)f_{R_{db,n}}(r,\theta_d){\rm d}r\bigg],\nonumber\\
P_{\rm cov,tb,\{l,n\}} &=\mathbb{E}_{\theta_d}\bigg[ \int_{0}^{\infty}\mathcal{A}_{tb,\{l,n\}}(d,\theta_d)P_{s,tb,\{l,n\}}(\tau,d)f_{R_{tb,\{l,n\}}}(r){\rm d}r\bigg],
%P_{\rm cov,tb_n} &=\mathbb{E}_{\theta_d}\bigg[\int_{0}^{\infty}\mathcal{A}_{tb,n}(d,\theta_d)P_{s,tb,n}(\tau,d)f_{R_{tb,n}}(r){\rm d}r\bigg],
\end{align}
in which $\theta_d$ is uniformly distributed in $(0,\frac{\pi}{2})$, the association probability $\mathcal{A}_{\{\cdot\}}$ is defined in (\ref{eq_associateprob}) in Lemma \ref{lemma_associateprob}, success probability $P_{s,\{\cdot\}}$ is defined in (\ref{eq_successprob}) in Lemma \ref{lemma_successprob}, and the distance distribution is defined in (\ref{eq_pdf}) in Lemma \ref{lemma_distancedistribution}.
\end{theorem}

\begin{remark}
		Observing from (\ref{eq_associateprob}) that UAVs are mainly associated with LoS TBSs and dedicated BSs since association probabilities are functions of CCDFs with horizontal distances $d_{\{\cdot\}}$. Taking $\mathcal{A}_{db,l}(d,\theta_d) = \bar{F}_{R_{db,n}}(d_{db,db,ln}(d),\theta_d)$ $ \bar{F}_{R_{tb,l}}(d_{db,tb,ll}(d))$ $ \bar{F}_{R_{tb,n}}(d_{db,tb,ln}(d))$ for example, the UAV associates with a LoS dedicated BS at a distance $d$ away and $d_{db,db,ln}(d)$ denotes the distance to the nearest NLoS dedicated BS and $d_{db,tb,ln}(d)$ denotes the nearest NLoS TBS. Generally, $\eta_n$ is much lower than $\eta_l$ and $\alpha_{ l}$ is less than $\alpha_n$, therefore, $d_{db,db,ln}(d) \approx d_{db,tb,ln}(d) \approx \Delta h_{\{1,2\}}$ which means that  $\bar{F}_{R_{db,n}}(d_{db,db,ln}(d),\theta_d) \approx \bar{F}_{R_{tb,n}}(d_{db,tb,ln}(d),\theta_d) \approx 1$. That is, UAVs have a very high probability of associating with LoS TBSs or dedicated BSs. Therefore, coverage probability is dominated by $P_{\rm cov,db,l}+P_{\rm cov,tb,l}$.
\end{remark}

\section{Trajectory Optimization}

In this section, we optimize the trajectory of a flying UAV to maintain a reliable connection with BSs. To do so, we first need to compute the mean SINR given the transmission distance, which is given in the following lemma.

\begin{lemma}[Mean SINR]
Given the transmission distance $d$, the averaged SINR conditioned on association with a dedicated BS or a TBS are, respectively,  given by
\begin{align}
	\label{eq_mean_SINR}
{\gamma}_{db}(d) =&\mathbb{E}_{\theta_d}\bigg[\sum_{c=\{l,n\}} P_{c}(d)\int_{0}^{\infty}\sum_{k = 1}^{m_c}\binom{m_c}{k}(-1)^{k+1}\mathcal{L}_{I_{db,c}+\sigma^2}(\xi_{db,c}(\tau,d),\theta_d){\rm d}\tau\bigg],\nonumber\\
{\gamma}_{tb}(d) =&\mathbb{E}_{\theta_d}\bigg[\sum_{c=\{l,n\}} P_{c}(d)\int_{0}^{\infty}\sum_{k = 1}^{m_c}\binom{m_c}{k}(-1)^{k+1}\mathcal{L}_{I_{tb,c}+\sigma^2}(\xi_{tb,c}(\tau,d)){\rm d}\tau\bigg].
\end{align}
\end{lemma} 
\begin{remark}
		We capture the correlation in the trajectory by considering the time correlation in the serving distance. For instance, we optimize the UAV trajectory while the locations of BSs are fixed and the distances to the serving BSs are correlated on time. For the interference correlation, authors in \cite{krishnan2017spatio} showed that the interference correlation coefficient decreases dramatically with slight increases in distances between two locations. Therefore, we investigate it in an average sense. For instance, we compute the mean SINR, which is a function of distance to the serving BSs while the interference, as well as channel fadings, are averaged.
Then, the trajectory optimization is based on the mean SINR.
\end{remark}

Now that the mean SINR is represented by a function of transmission distance, and we can start the trajectory optimization and to maximize $\bar{\gamma}$. Without loss of generality, we assume that the reference UAV travels from ${\rm S}(-L/2,0)$ to the destination located at ${\rm D}(L/2,0)$. This is due to the stationarity of  PPP and PLCP \cite{haenggi2012stochastic,dhillon2020poisson}. Given the initial location and destination, we first obtain the ground BSs that UAV associates with at the beginning and end of the trajectory, which are denoted by $\mathbf{g}_{s}$ and $\mathbf{g}_{d}$, which are both 2 by 1 vectors. Note that $\mathbf{g}_{s}$ and $\mathbf{g}_{d}$ are not necessarily to be the nearest BS to the ${\rm S}$ or the destination, e.g. $\mathbf{g}_{s}$ may not be the nearest BS to the ${\rm S}$, but the BS that provides the highest mean SINR. Besides, the maximal achievable mean SINR with $\mathbf{g}_{s}$ and $\mathbf{g}_{d}$ is an upper bound of the Max-Min SINR of the trajectory given an arbitrary but fixed realization, otherwise, the reference UAV cannot leave or reach the initial location or the destination. Let $\gamma_u$ denotes this upper bound. 

Let $\mathbf{g} = \{\mathbf{g}_1,\mathbf{g}_2,\cdots,\mathbf{g}_{n}\}$ be a sequence of the locations of ground BSs, either dedicated BSs or TBSs, that the reference UAV associates with along its trajectory, and $\mathbf{g}_1 = \mathbf{g}_s$ and  $\mathbf{g}_n = \mathbf{g}_d$. Note that $\mathbf{g}$ is a $2\times n$ matrix, where the first row denotes the $x$ coordinates and the second row denotes the $y$ coordinates. Let $\mathbf{r}(\gamma) = \{r_1(\gamma),\cdots,r_{n}(\gamma)\}$ be the sequence of the maximum transmission distance given a target mean SINR $\gamma$, in which $r_{\{\cdot\}}(\gamma)\in\{r_{tb}(\gamma),r_{db}(\gamma)\}$.
For a fixed realization, the constraint (\ref{eq_P1_SINR_constraint}) in $\mathcal{P}_2$ is satisfied if and only if a ground BS sequence existed which satisfies the following constraint,
\begin{align}
	\label{eq_constraint_dis}
	|\mathbf{g}_{i}-\mathbf{g}_{i-1}| \leq \mathbf{r}_{i}(\gamma)+\mathbf{r}_{i+1}(\gamma), \quad \forall i \in (2,n),
\end{align} 
therefore, the constraints for the locations of the association ground BSs are equivalent to,
\begin{align}
	|\mathbf{g}_{1}| &\leq r_{db}(\gamma) \quad\text{or}\quad r_{tb}(\gamma),\nonumber\\
	|\mathbf{g}_{i}-\mathbf{g}_{i-1}| &\leq \mathbf{r}_{i}(\gamma)+\mathbf{r}_{i+1}(\gamma), \quad \forall i \in (2,n),\nonumber\\
	|\mathbf{g}_{n}-{\rm D}(L,0)| &\leq r_{db}(\gamma) \quad\text{or}\quad r_{tb}(\gamma),
\end{align}
where $|\cdot|$ denotes the horizontal distance.

We now provide the details of solving $\mathcal{P}_1$ and obtaining the Max-Min SINR $\gamma^{*}$. In this step, we first construct an association BS graph based on the distances and $r_{tb}(\gamma)$ and $r_{db}(\gamma)$. If (\ref{eq_constraint_dis}) is satisfied, UAV can travel from these two BSs while maintaining a reliable connection. Hence, these two BSs are connected on the graph. After computing all the horizontal distances between the BSs, we obtain the graph showing the connection of BSs. The target SINR can be achieved if there exists at least one association BS sequence. Then, the Max-Min SINR is obtained by iterating over different values of $\gamma$.

\begin{figure}[ht]
	\centering
	\includegraphics[width=0.9\columnwidth]{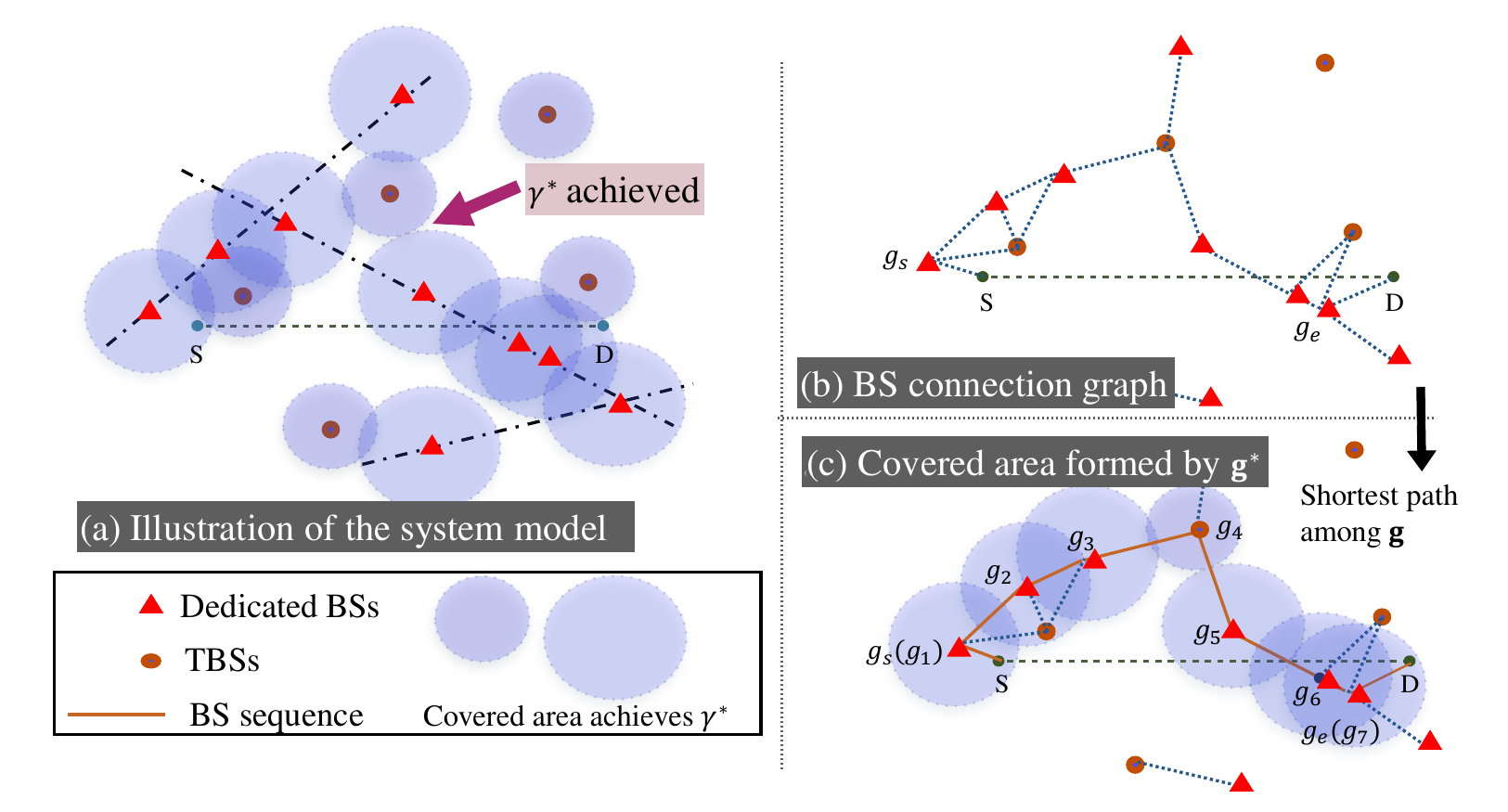}
	\caption{Illustration of \textbf{(a)} covered area formed by ground BSs that achieves  $\gamma^*$, \textbf{(b)} BS connection graph constructed by \textbf{(a)} and $\gamma^*$, \textbf{(c)} covered area formed by $\mathbf{g}^{*}$. }
	\label{fig_trajectory}
\end{figure}
We plot an illustration of the system model in Fig. \ref{fig_trajectory} in which $\gamma^*$ is achieved. In Fig. \ref{fig_trajectory}(a), the shadowed area is formed by the ground BSs with radii $r_{tb}(\gamma^*)$ and $r_{db}(\gamma^*)$. If the UAV flies within the shadowed area,  $\gamma^t\geq\gamma^*$ at any time $t$ can always be achieved, e.g., the minimum $\gamma^t = \gamma^*$ achieved at the boundary. Hence, if we can find a trajectory from ${\rm S}$ to ${\rm D}$ within the covered area, the target SINR can be satisfied.
In Fig. \ref{fig_trajectory}(b), BSs are connected if (\ref{eq_constraint_dis}) is satisfied, and as shown, some BS association sequences can be found to connect ${\rm S}$ and ${\rm D}$. Among all the BS association sequence, let $\mathbf{g}^*$ denote the BS sequence that has the shortest distance, which is shown in Fig. \ref{fig_trajectory} (c).
%To solve $\mathcal{P}_2$, we use a similar method as mentioned in \cite{}, but slightly different since we have a large-scale network, hence, obtaining all the possible association BS sequence is tricky, and our goal is to obtain ${\rm SINR^{*}}$ and no constraint on the mission complete time. The proposed method has two steps: (i) we obtain ${\rm SINR^{*}}$ by the proposed Algorithm \ref{Alg_MaxMinSINR} and the Dijkstra algorithm is used to obtain an approximation solution, which achieves the same ${\rm SINR^{*}}$ but has a slightly longer minimal time path, (ii) we compute the minimal time path given the possible associated BS sequence obtained by Dijkstra algorithm.

%To obtain the locations of ground BS sequence, we use the Dijkstra algorithm and maximize the minimal achievable averaged SINR of the UAV trajectory, we propose the Algorithm \ref{Alg_MaxMinSINR}.
% Besides, since we use stochastic geometry tools to model the large-scale deployment of ground BSs, it is impossible  
 Due to the nature of large-scale deployment of BSs, which results in a huge number of possible association BS sequences, we choose the the BS sequence that has the shortest horizontal distance, which is obtained by Dijkstra algorithm and denoted by $\mathbf{g}^{*}$, and optimize the UAV trajectory based on $\mathbf{g}^{*}$. The algorithm for the first step is proposed in Algorithm \ref{Alg_MaxMinSINR}. The complexity of the proposed algorithm is $\mathcal{O}(\bar{M}!)$, where $\bar{M} = (\lambda_{tb}+\lambda_{db})S$ denotes the average number of BSs in the simulation area and $S$ is the simulation area.

\begin{algorithm}
	\caption{Algorithm for Max Min SINR and $\mathbf{g}^{*}$}
	\SetAlgoLined
	\DontPrintSemicolon
	\KwIn{$\Phi_{\{tb,db\}}$: Locations of the ground BSs\newline
		${\rm S},{\rm D}$: Locations of the inital point and destination}
	\KwOut{$\gamma^*$: Max Min SINR, $\mathbf{g}$, $\mathbf{r}(\gamma^{*})$ }
	\textbf{Initialization:} $s_m=\gamma_{m}$, $s_u=\gamma_{u}$, $\mathbf{g}$, $\mathbf{r}$, $i = 0$, $s^{0}=s_m$ \newline
	\SetKwFunction{FMain}{MaxMinSINR}
	\SetKwProg{Fn}{Function}{:}{}
	\Fn{\FMain{$\Phi_{\{tb,db\}}$ ,${\rm S}$ ,${\rm D}$}}{
		\Repeat{$|s^{i}-s^{i-1}|<\epsilon$}{
		Update $r_1 = r_{tb}(s^{i})$, $r_2 = r_{db}(s^{i})$	\\
		Compute the horizontal distance between any two ground BSs, if the distance is shorter than the sum of their maximal transmission distances, either $r_1$ or $r_2$, save the value, otherwise, save as infinity\\
		Construct the graph based on the obtained distance and find the shortest path between $g_s$ and $g_d$ via Dijkstra algorithm\\
		\eIf{If the distance of the shortest path is finite}{
		Update $s^{i+1} = \frac{s_u+s^{i}}{2}$, $s_{m} = s^{i}$, $\mathbf{g}$, $\mathbf{r}( \gamma^{*})$\
		}{
		Update $s^{i+1} = \frac{s_m+s^{i}}{2}$, $s_{u} = s^{i}$
		}
		Update $i = i+1$
		}
	Update $\gamma^* = s^{i}$\\
		\textbf{return} $\gamma^*$, $\mathbf{g}^{*}$, $\mathbf{r}(s^{i})$
	}
	\textbf{End Function}
	\label{Alg_MaxMinSINR}
\end{algorithm}

In what follows, we optimize the UAV trajectory to minimize the traveling time $T_{\rm min} $ based on the association BS sequence $\mathbf{g}^{*}$. We obtain the optimal trajectory  by exhaustive search which is similar to Method II introduced in \cite{8531711}, but we modify this method by adding one more step to achieve a better performance. However, this trajectory is still a sub-optimal solution and slightly worse than the real shortest time path, denoted by $\mathbf{u}^*(t)$, which is obtained by Method I in \cite{8531711}, since we only consider a route within the covered area formed by the BS sequence $\mathbf{g}^*$. 	The reason that the Method I in \cite{8531711} is not practical in this work is that we consider a large-scale network and a large number of   BSs, which enables us to analyze the interference but cause an infinite number of possible BS association sequences. Therefore, obtaining all the possible BS association sequences and all the possible paths that can achieve the Max-Min SINR is tricky.

\begin{figure}[ht]
	\centering
	\includegraphics[width=0.9\columnwidth]{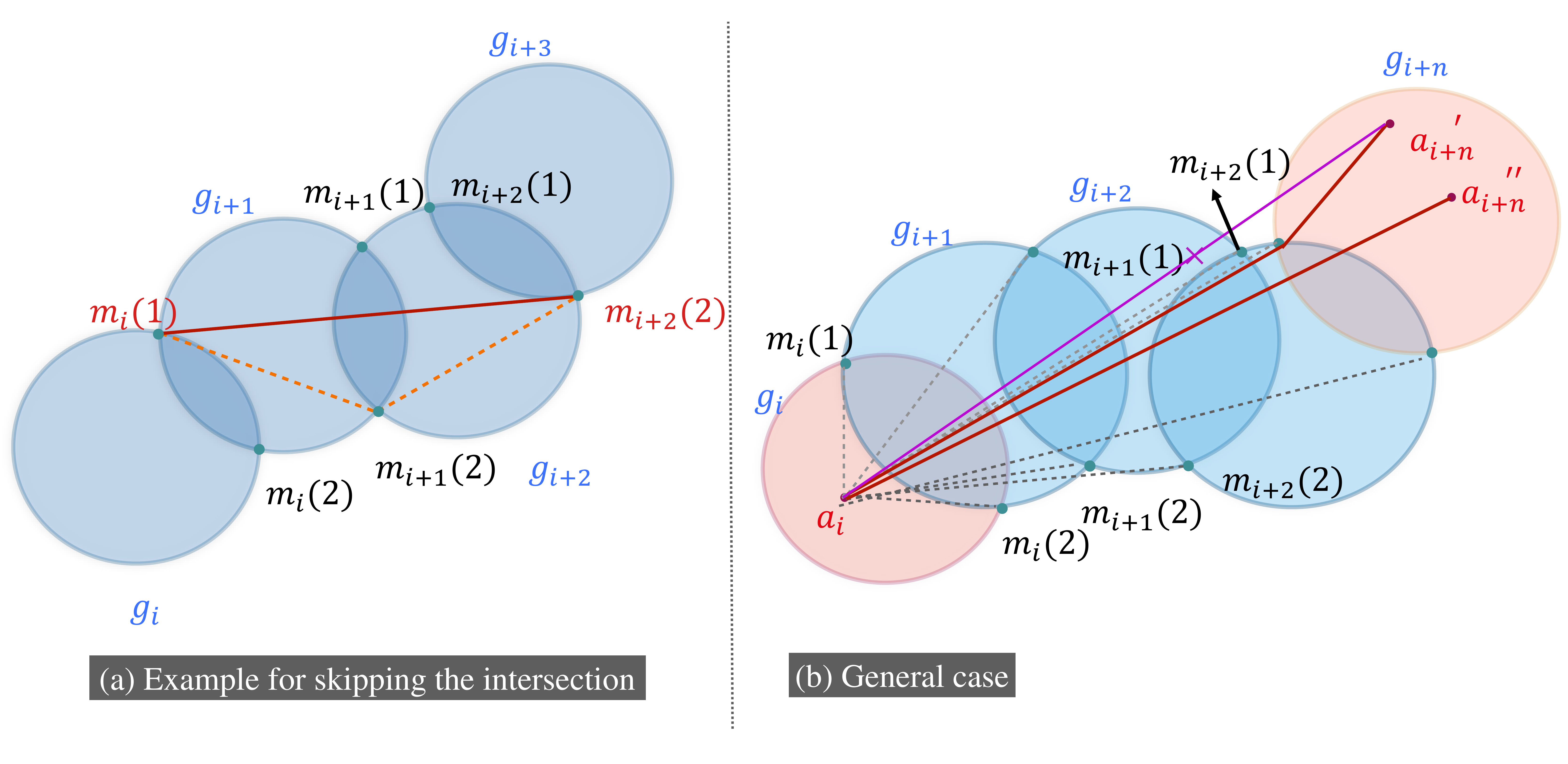}
	\caption{Illustration of Lemma \ref{lemma_segment}. \textbf{(a)} Instead of traveling from $\mathbf{m}_{i}(1)$ to $\mathbf{m}_{i+1}(2)$ to $\mathbf{m}_{i+2}(2)$, the UAV can fly from $\mathbf{m}_{i}(1)$ to $\mathbf{m}_{i+2}(2)$ directly. \textbf{(b)} General case.}
	\label{fig_trajectory2}
\end{figure}

As mentioned in \cite{8531711} Method II, an approximation solution of the UAV trajectory is obtained by graph constructed by the intersections of the circles formed by the ground BSs and their maximal transmission distances.  Observing that UAVs are not necessarily to travel to the intersections of each BS one by one, it can skip some intersections. For instance, in Fig. \ref{fig_trajectory2} (b), if the reference UAV travels from $\mathbf{a}_i$ to  $\mathbf{a}_{i+n}^{''}$, it can travel along a straight line without passing though any intersections. It implies that if we plot a connection graph, the points $\mathbf{a}_i$ and $\mathbf{a}_{i+n}^{''}$ should be connected directly, and the points $\mathbf{a}_i$ and  $\mathbf{a}_{i+n}^{'}$ should  not be connected since its segment is not totally covered. Therefore, instead of boundary quantization, we use the following lemma to check the connection between intersections, ${\rm S}$ and ${\rm D}$.

\begin{lemma}[Possible Traveling Path of UAVs]
	\label{lemma_segment}
Let $\mathbf{m}_{i}(1), \mathbf{m}_{i}(2)$ be the locations of two intersections of $B(\mathbf{g}_{i},r_{i})$ and $B(\mathbf{g}_{i+1},r_{i+1})$, and  $\mathbf{m}_{i+1}(1), \mathbf{m}_{i+1}(2)$ be the locations of two intersections of $B(\mathbf{g}_{i+1},r_{i+1})$ and $B(\mathbf{g}_{i+2},r_{i+2})$, as shown in Fig. \ref{fig_trajectory2} (a). If the segment $\overrightarrow{\mathbf{m}_{i}(1)\mathbf{m}_{i+1}(1)}$ is covered by the $B(\mathbf{g}_{i},r_{i})$ and $B(\mathbf{g}_{i+1},r_{i+1})$, it should satisfy the following constraints,
\begin{align}
k_{\mathbf{m}_{i}(1)\mathbf{m}_{i+1}(1)} \geq k_{\mathbf{m}_{i}(1)\mathbf{m}_{i+2}(1)} \geq k_{\mathbf{m}_{i}(1)\mathbf{m}_{i+1}(2)},
\end{align}
where $k_{ab}$ denotes the slope of the segment $\overrightarrow{ab}$. Consequently,  for a general case, if a path $\overrightarrow{\mathbf{a}_{i}\mathbf{a}_{i+n}}$ is covered by the circles formed by the BS $\mathbf{g}_{i},\cdots,\mathbf{g}_{i+n}$, it should satisfy,
\begin{align}
k_{\mathbf{a}_{i}\mathbf{a}_{i+n}} &\leq \min(k_{\mathbf{a}_{i}\mathbf{m}_{i+n}(1)}, k_{\mathbf{a}_{i}\mathbf{m}_{i+n-1}(1)} \cdots,  k_{\mathbf{a}_{i}\mathbf{m}_{i}(1)}),\nonumber\\
k_{\mathbf{a}_{i}\mathbf{a}_{i+n}} &\geq \max(k_{\mathbf{a}_{i}\mathbf{m}_{i+n}(2)}, k_{\mathbf{a}_{i}\mathbf{m}_{i+n-1}(2)} \cdots, k_{\mathbf{a}_{i}\mathbf{m}_{i}(2)}).
\end{align}
\end{lemma}

\begin{IEEEproof}
	The geometric illustration is shown in Fig. \ref{fig_trajectory2} (b):  $\overrightarrow{\mathbf{a}_{i}\mathbf{a}_{i+n}^{''}}$ is covered by the circles since $k_{\mathbf{a}_{i}\mathbf{a}_{i+n}^{''}}$ is in the range, $\overrightarrow{\mathbf{a}_{i}\mathbf{a}_{i+n}^{'}}$ satisfies the constraints due to its slope is greater than $\min(k_{\mathbf{a}_{i}\mathbf{m}_{i+n}(1)}, k_{\mathbf{a}_{i}\mathbf{m}_{i+n-1}(1)} \cdots,  k_{\mathbf{a}_{i}\mathbf{m}_{i}(1)})$.
\end{IEEEproof}
\begin{remark}
		This step is a simplification of the discrete boundary step in \cite{8531711} method II. If the segment of the initial location and the destination is covered, say $\mathbf{a}_{i},\mathbf{a}_{i+n}^{''}$, then the minimal distance trajectory is achieved by the line $\overrightarrow{\mathbf{a}_{i}\mathbf{a}_{i+n}^{''}}$. In this case, discretizing the boundary over-complicates the analysis since we do not compute the handover locations. In the case that the initial location and the destination are not covered, $\mathbf{a}_{i},\mathbf{a}_{i+n}^{'}$, the minimal distance trajectory is achieved at the boundary point, e.g., $|\overrightarrow{\mathbf{a}_{i}\mathbf{m}_{i+2}(1)}|+|\overrightarrow{\mathbf{m}_{i+2}(1)\mathbf{a}_{i+n}^{'}}|$.
\end{remark}

Now we are able to construct a graph which is composed of all the possible paths that the reference UAV can travel within the area covered by the circles formed by BS sequence $\mathbf{g}^{*}$. The algorithm for this step is provided in Algorithm \ref{Alg_Tmin}.

\begin{algorithm}
	\caption{Algorithm for minimal time trajectory}
	\SetAlgoLined
	\DontPrintSemicolon
	\KwIn{$\mathbf{g}^{*}$: Sequence of the locations of the ground BSs\newline
		 $\mathbf{r}(\gamma^{*})$: Sequence of the maximal transmission distance}
	\KwOut{$T_{\rm min}$: Minimal traveling time, $\mathbf{u}$: UAV trajectory}
	\textbf{Initialization:} $\mathbf{u}$, $T_{\rm min}$ \newline
	\SetKwFunction{FMain}{MinTime}
	\SetKwProg{Fn}{Function}{:}{}
	\Fn{\FMain{$\mathbf{g}^{*}$, $\mathbf{r}({\rm \gamma^{*}})$, S, D}}{
		Compute the intersections given the BS sequence $\mathbf{g}^{*}$ and the radii $\mathbf{r}(\gamma^{*})$\\
		Use Lemma \ref{lemma_segment} to obtain all the segments between the intersections, ${\rm S}$ and ${\rm D}$, which are covered by the area formed by $\mathbf{g}^{*}$ with radii $\mathbf{r}(\gamma^{*})$\\
		Compute the shortest path $\mathbf{u}$ and the minimal time $T_{\rm min}$ via Dijkstra algorithm
		\textbf{return} $\mathbf{u}$, $T_{\rm min}$
	}
	\textbf{End Function}
	\label{Alg_Tmin}
\end{algorithm}

\begin{figure}[ht]
	\centering
	\includegraphics[width=0.9\columnwidth]{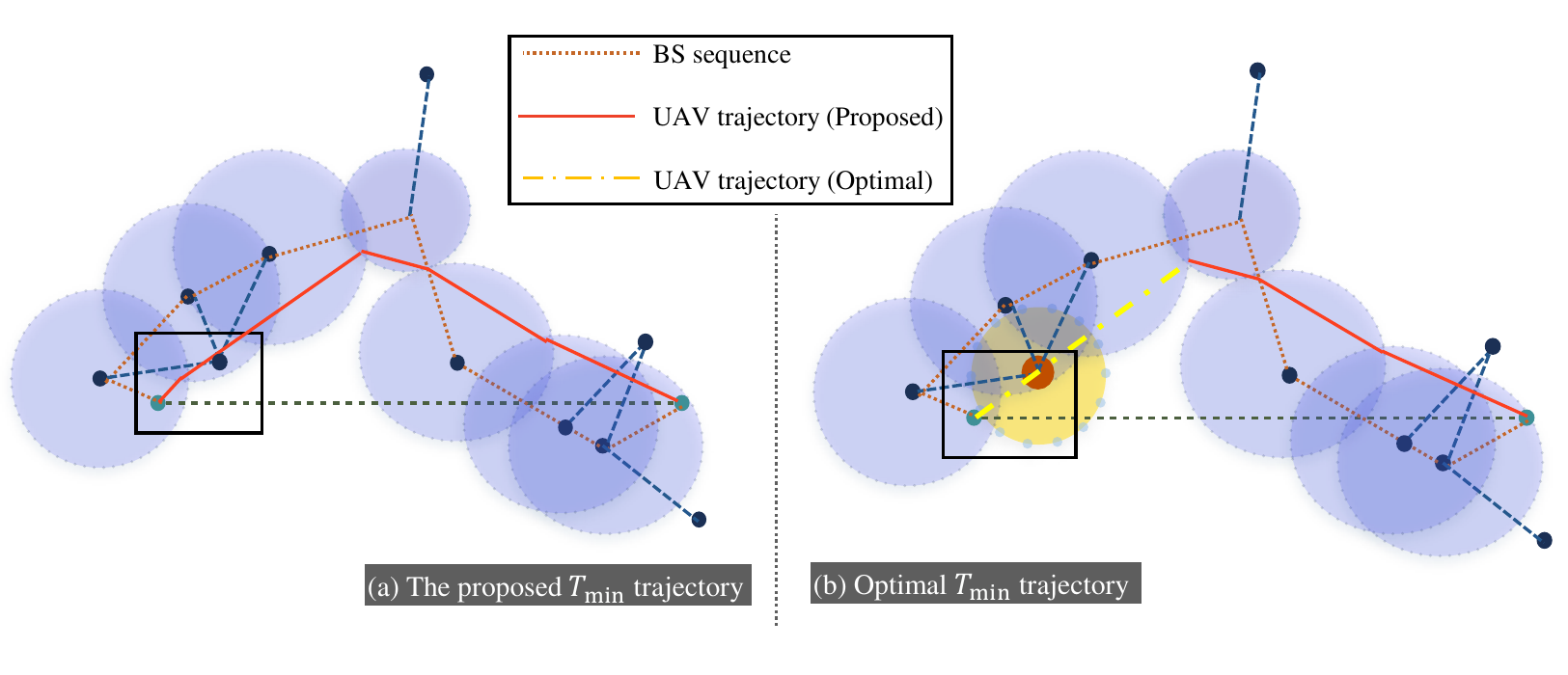}
	\caption{The proposed trajectory which is a sub-optimal solution to the minimal time trajectory.}
	\label{fig_trajectory3}
\end{figure}

However, we would like to point out that the $T_{\rm min}$ obtained by Algorithm \ref{Alg_Tmin} is a sub-optimal solution even though we use exhaustive search since we only consider the routes among the BS sequence $\mathbf{g}^*$, but this do not influence the obtained Max-Min SINR since $\gamma^*$ is obtained by considering all the BSs. For instance, in Fig. \ref{fig_trajectory3}(a) we plot the $\mathbf{u}(t)$ obtained by Algorithm \ref{Alg_Tmin}. Since we only consider the covered area formed by $\mathbf{g}^*$, which has a smaller covered area than considering all the BSs, such as the yellow area in Fig. \ref{fig_trajectory3}(b), the obtained $\mathbf{u}(t)$ is slightly longer than the optimal solution, e.g., in Fig. \ref{fig_trajectory3}(a) the reference UAV travels to the intersection while it travels in a straight line in Fig. \ref{fig_trajectory3}(b).

\section{Numerical Results}
In this section, we validate our analytical results with simulations and evaluate the impact of various system parameters on the network performance. Unless stated otherwise, we use the simulation parameters as listed herein Table \ref{par_val}.
\begin{table}\caption{Table of Parameters}\label{par_val}
	\centering
	\begin{center}
		\resizebox{0.75\columnwidth}{!}{
			\renewcommand{\arraystretch}{1}
			\begin{tabular}{ {c} | {c} | {c}  }
				\hline
				\hline
				\textbf{Parameter} & \textbf{Symbol} & \textbf{Simulation Value}  \\ \hline
				Density of TBSs & $\lambda_{ tb}$ & 1 km$^{-2}$ \\ \hline
				Height of TBSs, dedicated BSs and altitude of UAVs & $h_{tb}$, $h_{db}$, $h_u$ & $30$, $10$, $100$ m \\ \hline
				Transmission power of TBS and dedicated BSs  & $\rho_{tb}$, $\rho_{db}$ & $1$,  $1$  w\\ \hline
				Mainlobe, sidelobe gain and coverage range & $g_m$, $g_s$, $z_{db}$ & $10$,  $0$ dB, $534$ m\\ \hline
				UAV traveling distance and velocity & $L$, $v$ & $5$  km, $18$ m/s\\ \hline
				SINR threshold & $\tau$ & $0$ dB\\ \hline
				N/LoS environment variable & $a, b$ & 12, 0.11 \\\hline
				Noise power & $\sigma^2 $ & $10^{-9}$ W\\\hline
				N/LoS and TBS path-loss exponent & $\alpha_{ n},\alpha_{ l},\alpha_{ tb}$ & $4,2.1,4$ \\\hline
				N/LoS fading gain & $m_{ n},m_{ l}$ & $1,3$ \\\hline
				N/LoS additional loss& $\eta_{ n},\eta_{ l}$ & $-20,0$ dB 
				\\\hline\hline
		\end{tabular}}
	\end{center}
	%\vspace{-8mm}
\end{table}

\begin{figure}
	\centering
	\subfigure[]{\includegraphics[width=0.7\columnwidth]{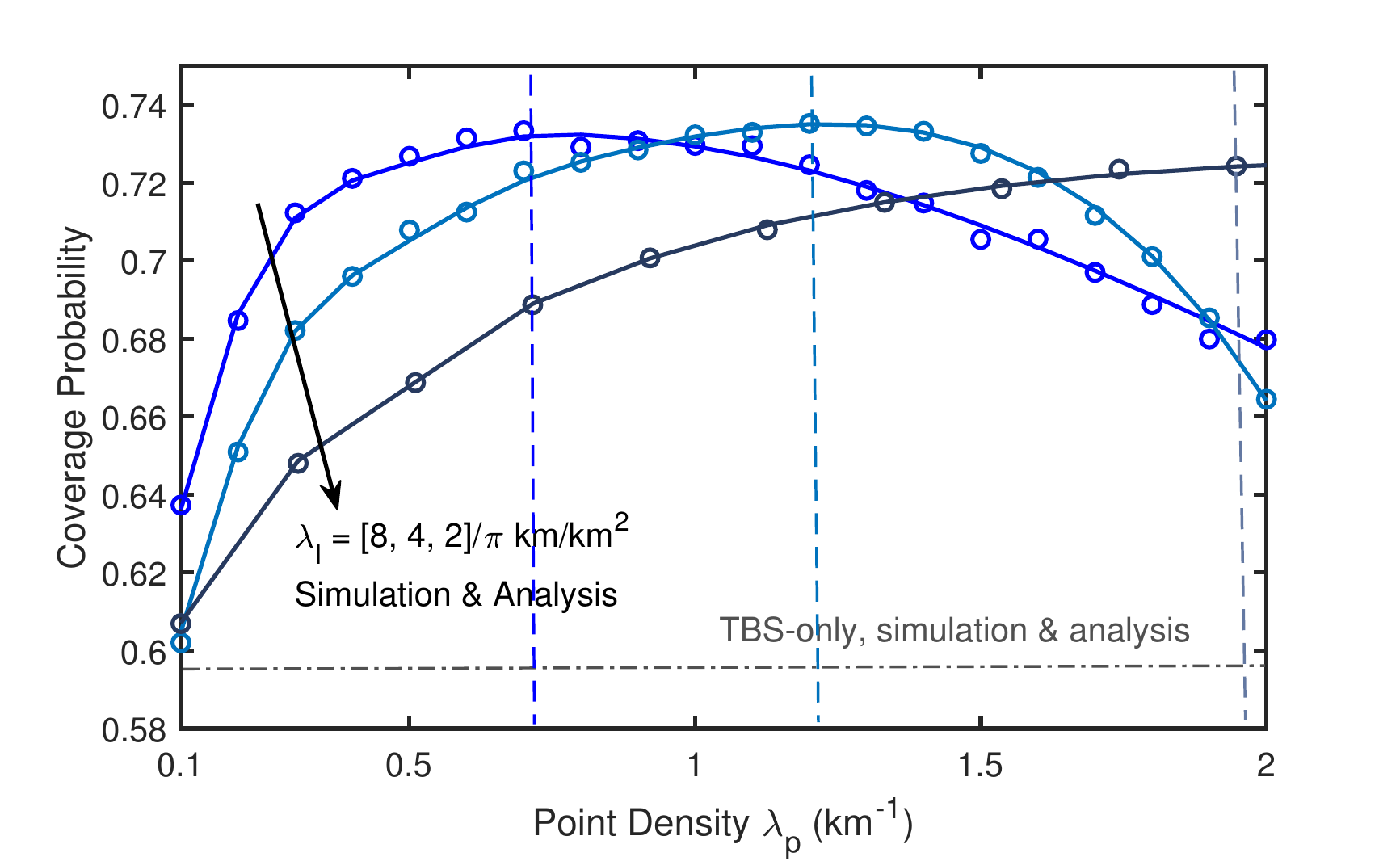}}
	\subfigure[]{\includegraphics[width=0.7\columnwidth]{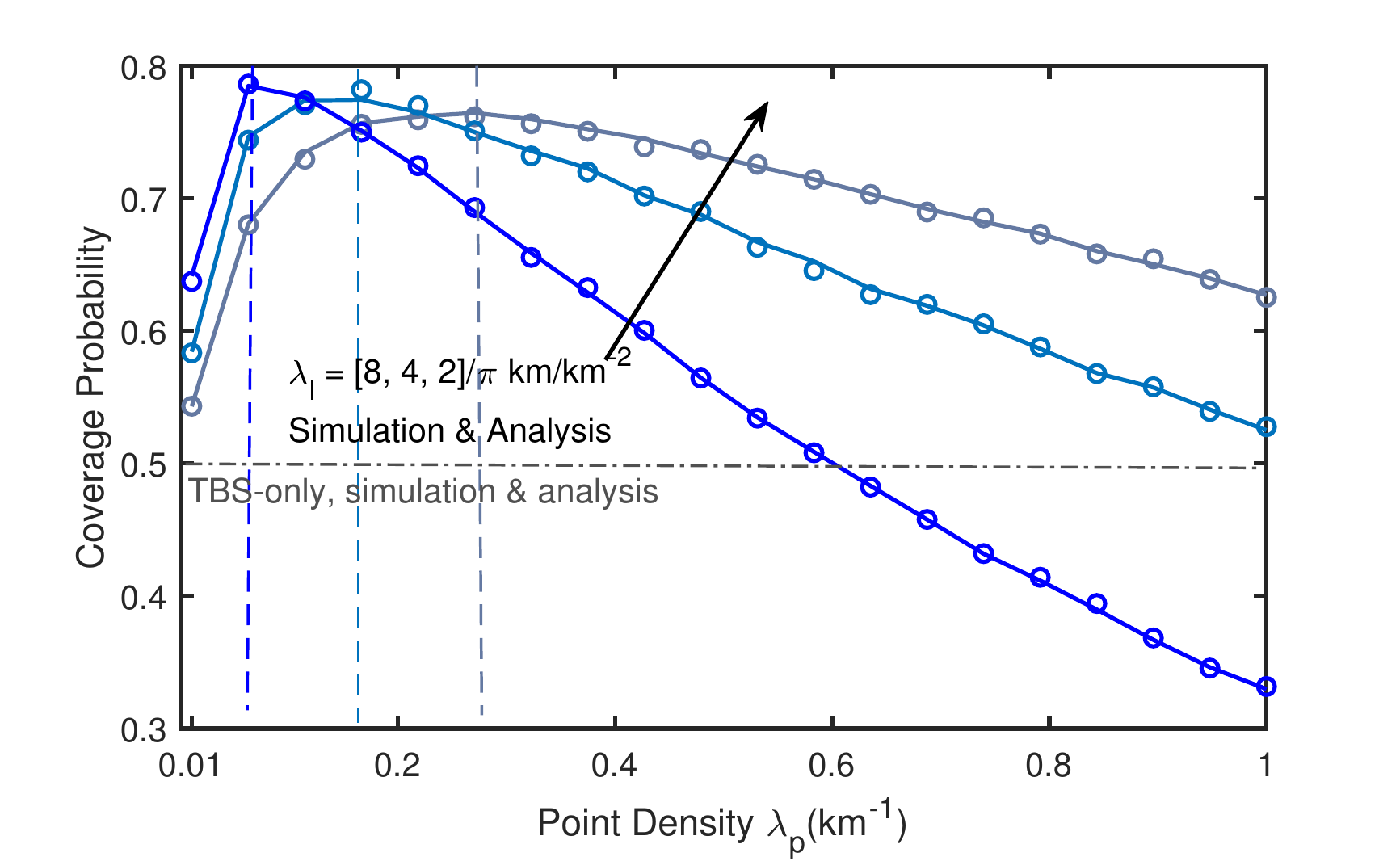}}
	\caption{Coverage probability in the case of \textbf{(a)} high dense urban area ($a = 12, b = 0.11$)  and $\lambda_{tb} = 1$ km$^{-2}$, \textbf{(b)} rural area ($a = 4.88, b = 0.43$, $\eta_{\{l,n\}} = 0$ dB) and $\lambda_{tb} = 0.1$ /km$^{-2}$.}
	\label{fig_coverageprob}
\end{figure}

For the simulation of the considered system setup, we compute the coverage probability and the Max-Min SINR separately and run a large number of iterations, $10^3$ for Max-Min SINR and $10^6$ for coverage probability, to ensure accuracy.  For each iteration of computing coverage and Max-Min SINR, we first generate a PPP realization for the locations of TBSs and a PLCP realization for the locations of the dedicated BSs. For the coverage probability simulation, assuming the typical UAV is located at the origin, we compute the SINR at the origin. Besides, channel fading is also generated for each iteration. For the simulation of the Max-Min SINR, we first compute the SINR given the transmission distance in the case of different road and point densities. Next, we use   Algorithm \ref{Alg_MaxMinSINR} and Algorithm \ref{Alg_Tmin} to compute the $\gamma^{*}$ and $T_{\rm min}$. Note that the randomness of BS locations is considered since PPP and PLCP realizations vary with iterations.
To avoid UAVs traveling a very long distance, we define a simulation area $S = 25\times 25$ km$^2$ to restrict the traveling area of UAVs. For the road density, we choose the value $\lambda_l = (2,4,8)/\pi$ km/km$^2$ in which low value is used for rural areas and high value is used for urban areas \cite{lu2022transport}.

In Fig. \ref{fig_coverageprob}, we plot the coverage probability of the considered network. In high dense urban area, with the increase of the dedicated BS density, the coverage probability increases first due to a better communication channel and higher received signal power. However, with further increase of the dedicated BS density, the coverage probability decreases, which is because of higher interference. Our results show that optimal value of point density exists for different values of road density. With the decrease of the road density, the optimal value of point density increases and the optimal dedicated BS density is about $\lambda_l\lambda_p\pi \approx 4$ km$^{-2}$. In rural areas (in which $\lambda_{tb} = 0.1$ km$^{-2}$), the network also benefits from deploying dedicated BSs but the density of dedicated BSs is much lower than high dense urban area, which is about $\lambda_l\lambda_p\pi \approx 0.75$ km$^{-2}$. Optimal dedicated BS density decreases with the increases of the probability of establishing LoS links due to the interference. We would like to point out that even if the dedicated BSs are modeled by a PPP, aerial users coverage performance would still improve. However, as stated in the motivation, it is more practical to model them using PLCP to accurately capture the scenario of deploying them on roadside furniture, which we believe is a most more cost and space efficient deployment strategy.

\begin{figure}
	\centering
	\subfigure[]{\includegraphics[width=0.49\columnwidth]{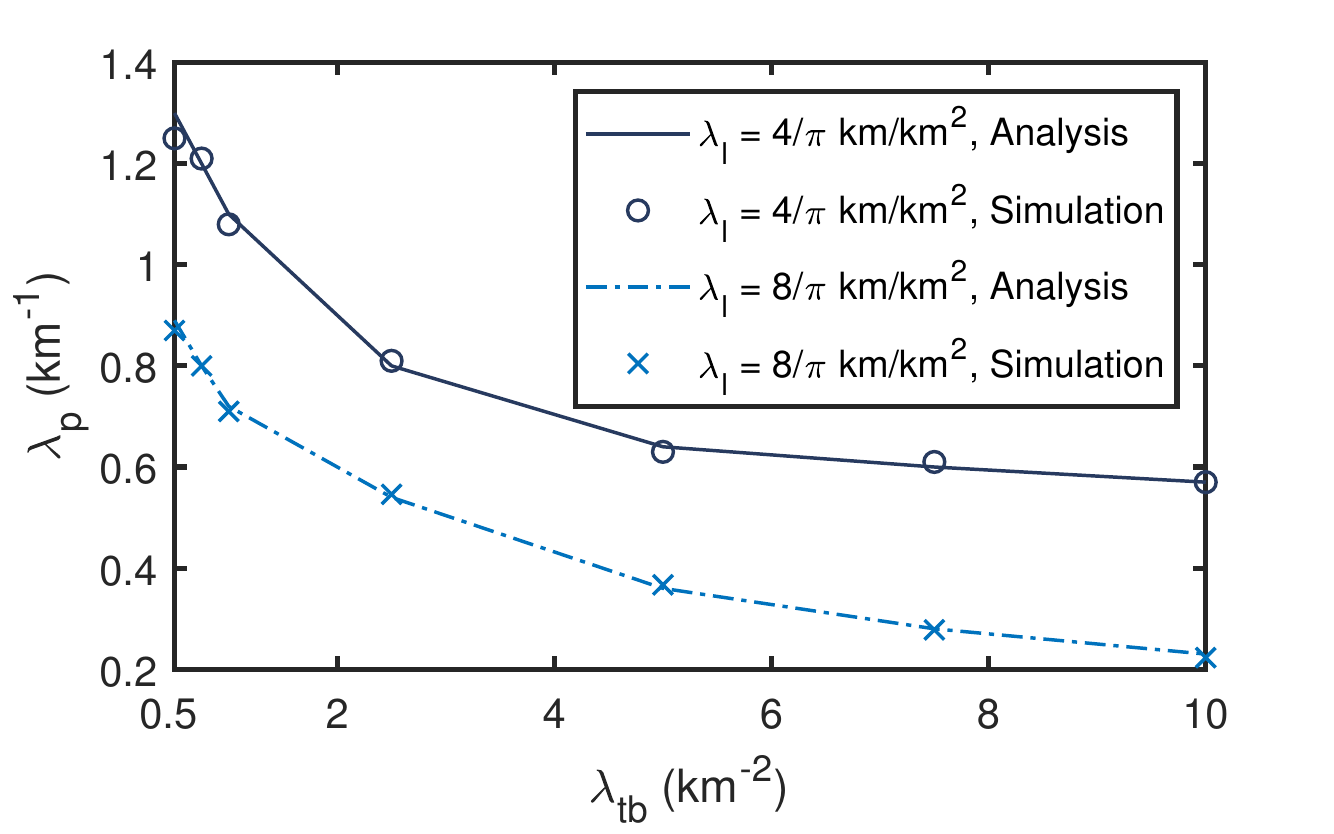}}
	\subfigure[]{\includegraphics[width=0.49\columnwidth]{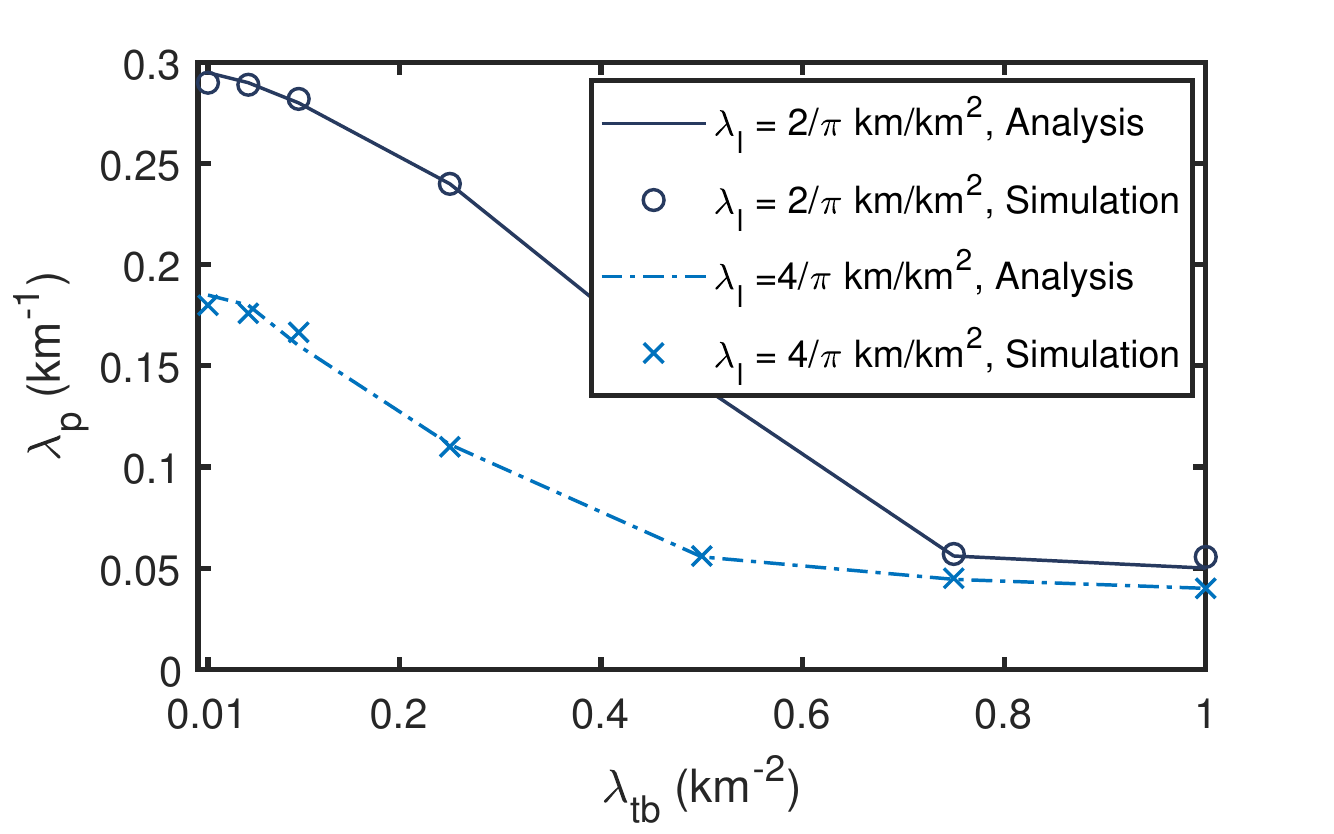}}
	\caption{Optimal point density under different TBS density in the case of \textbf{(a)} high dense urban area ($a = 12, b = 0.11$), \textbf{(b)} rural area ($a = 4.88, b = 0.43$, $\eta_{\{l,n\}} = 0$ dB).}
	\label{fig_coverageprobdiflambdatb}
\end{figure}
To further study the influence of TBS density on the dedicated BS density, we plot optimal $\lambda_p$ under different $\lambda_{tb}$ and road density in Fig. \ref{fig_coverageprobdiflambdatb}, in which optimal $\lambda_p$ is obtained by maximizing the coverage probability. We show that optimal $\lambda_p$ decreases with the increase of $\lambda_{tb}$ and rural areas require less dedicated BSs compared Fig. \ref{fig_coverageprobdiflambdatb} (a) with Fig. \ref{fig_coverageprobdiflambdatb} (b).

\begin{figure}
	\centering
	\includegraphics[width=0.8\columnwidth]{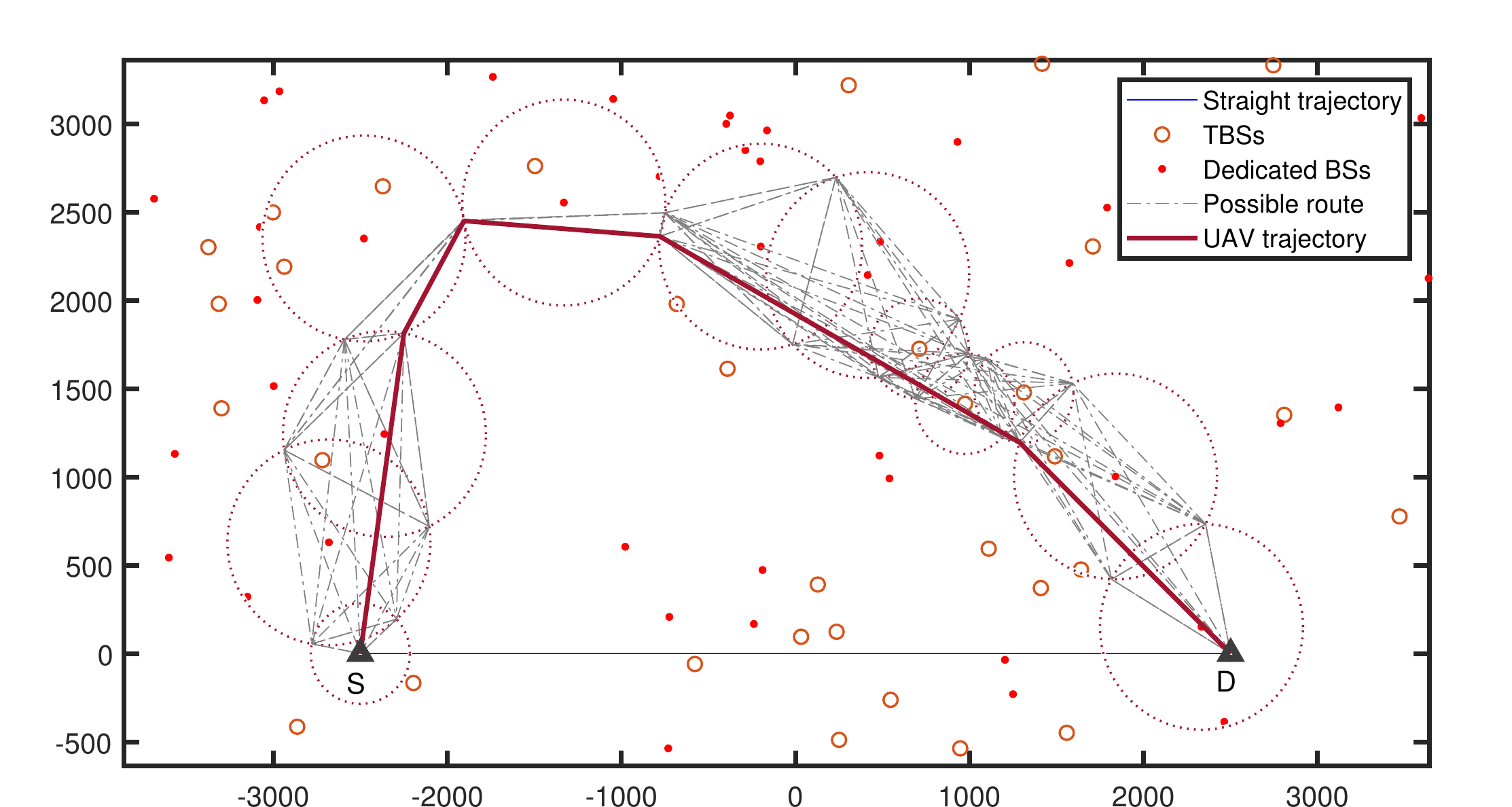}
	\caption{Illustration of minimal time UAV trajectory which maximizes the minimal SINR when $\lambda_l = 4/\pi$ km/km$^2$ and $\lambda_p = 0.4$ km$^{-1}$. }
	\label{fig_illustration}
\end{figure}

Fig. \ref{fig_illustration} shows the final minimal time trajectory of the reference UAV when $\lambda_l = 4/\pi$ km/km$^2$ and $\lambda_p = 0.4$ /km and the red dash circle denotes the maximal transmission distance of the BSs that achieves $\gamma^* = 17.8$ dB. Since we consider that the BS on the street furniture is dedicated to aerial users, the maximal transmission distance of dedicated BSs is much larger than traditional TBSs. Besides, this trajectory shows the use of Lemma \ref{lemma_segment} to skip some intersections, hence, achieving a better performance without exhaustive search among the whole covered area.

\begin{figure}[ht]
		\centering
\subfigure[]{\includegraphics[width=0.7\columnwidth]{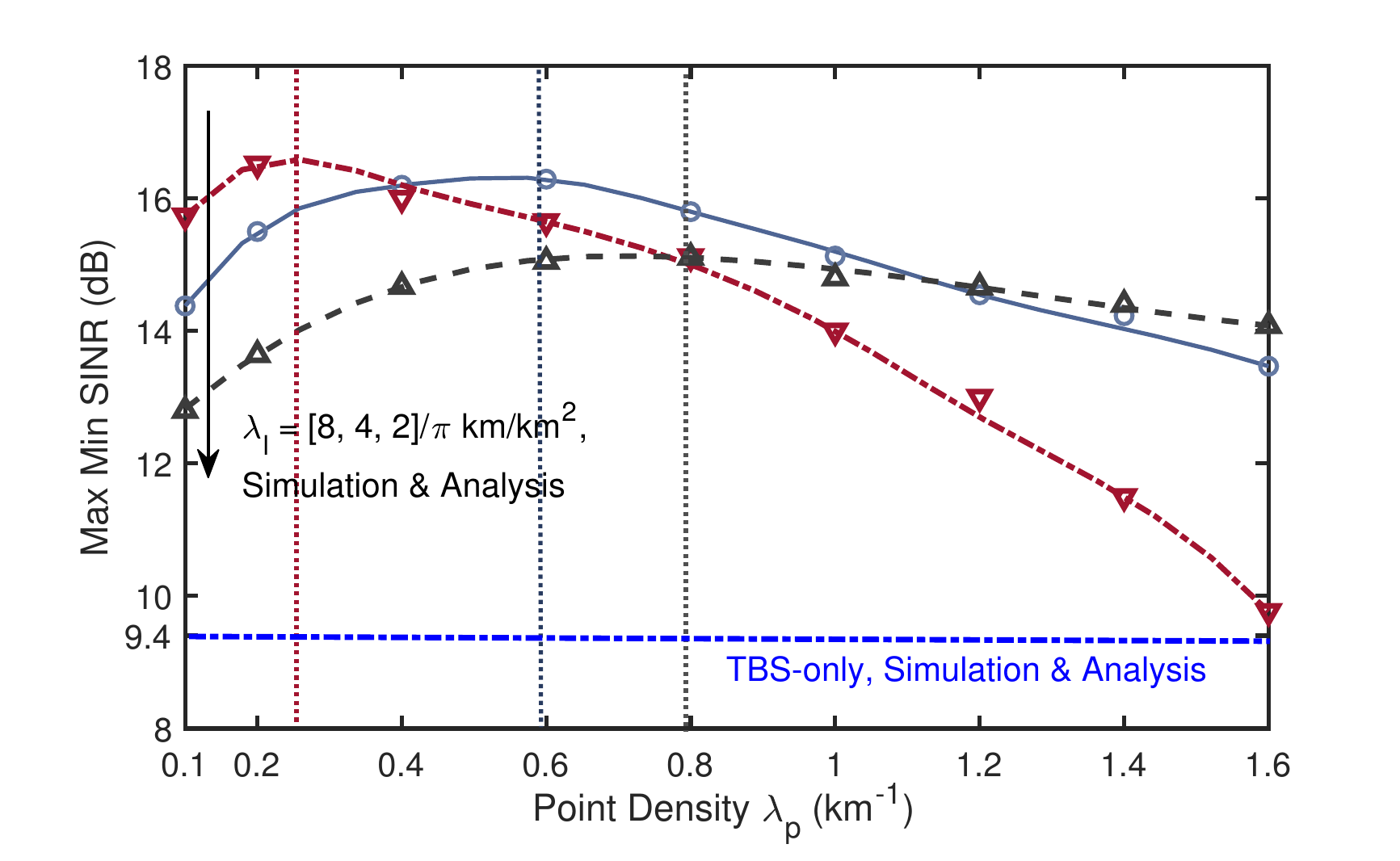}}
\subfigure[]{\includegraphics[width=0.7\columnwidth]{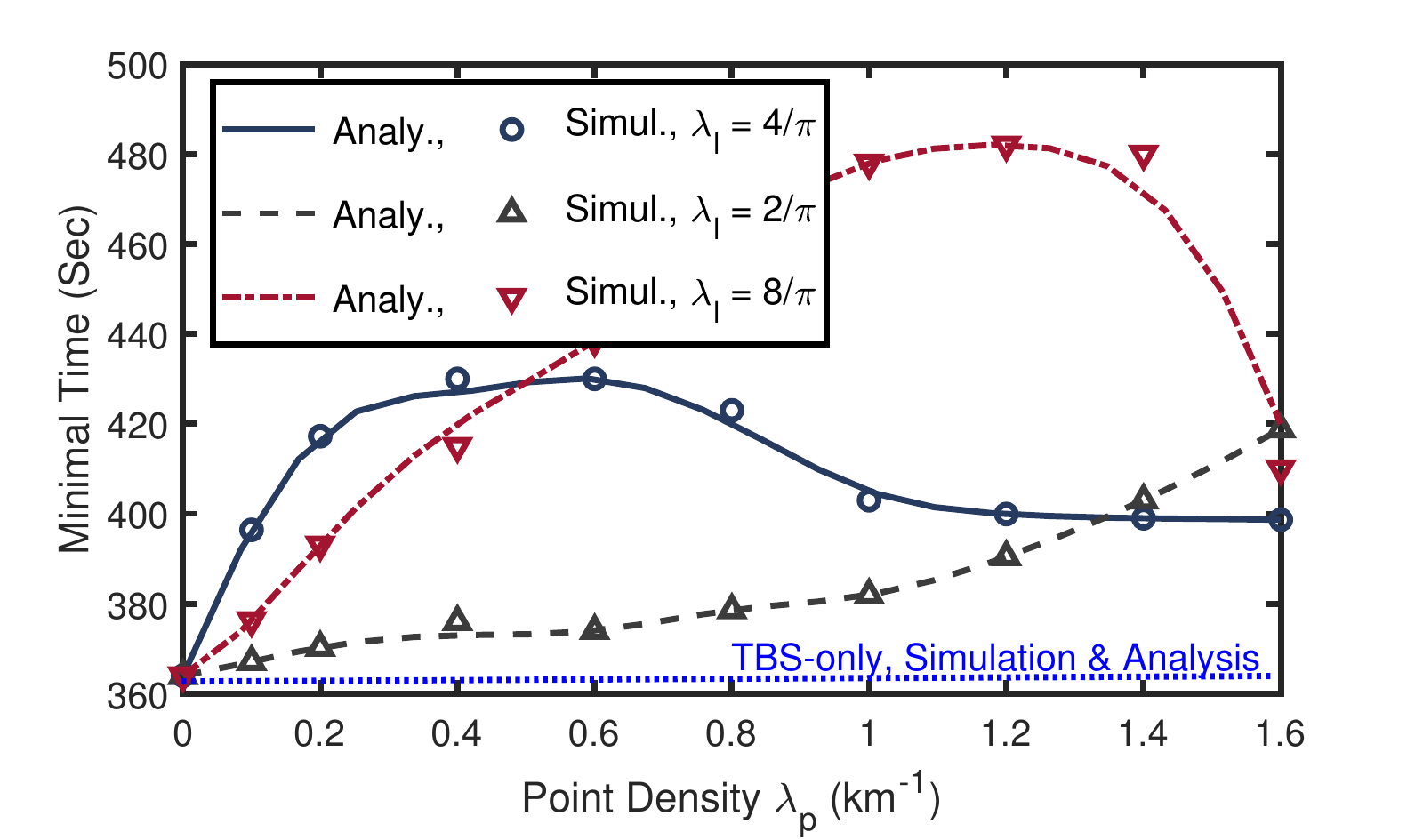}}
	\caption{\textbf{(a)} Max-Min SINR under different line densities and dedicated BS density (point density). \textbf{(b)} $T_{\rm min}$ of the UAV trajectory achieves $\gamma^*$ and $\lambda_l$ is in km/km$^{2}$.}
	\label{MaxMinSINR}
\end{figure}
Fig. \ref{MaxMinSINR} shows the results of UAV trajectory optimization, in which the analysis results are obtained based on (\ref{eq_mean_SINR}) and the two proposed algorithms. 
In Fig. \ref{MaxMinSINR} (a), we show the Max-Min SINR under different road densities. While the deployment of dedicated BSs improves the Max-Min SINR compared with the TBS-only network, we show the optimal point densities for different line densities and have the same shift trend as Fig. \ref{fig_coverageprob} (a). For a given line density, with the increase of the point densities, the distance between BSs decreases, however, the mean SINR given a transmission distance also decreases due to higher interference. Therefore, both sides in (\ref{eq_constraint_dis}) decreases. At low point density, the distances decrease dramatically, while at high point density the decrease of $\gamma$ has a higher impact.

\begin{figure}
	\centering
	\includegraphics[width=0.9\columnwidth]{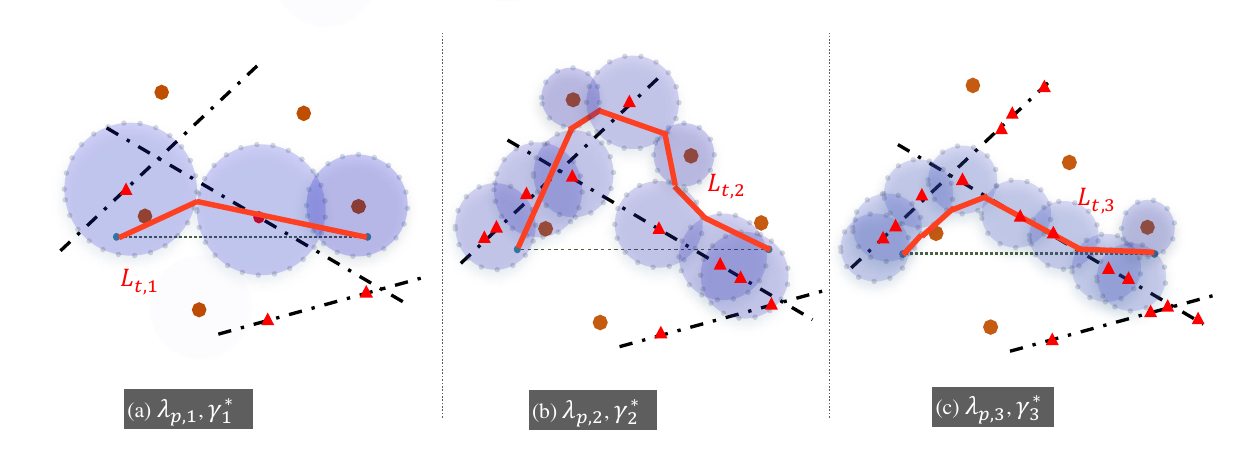}
	\caption{UAV trajectories in three different point densities, $\lambda_{p,1}<\lambda_{p,2}<\lambda_{p,3}$, and achieve three different Max-Min SINR $\gamma_{1}^{*}<\gamma_{3}^{*}<\gamma_{2}^{*}$, while the road density keeps the same.}
	\label{fig_trajectory4}
\end{figure}

In Fig. \ref{MaxMinSINR} (b), we show the $T_{\rm min}$ under different road densities. Interestingly, $T_{\rm min}$ increases first with the increase of point density and then decreases with the further increase of $\lambda_p$. The reasons are as follows. At low point densities (low BS density), ${\rm S}$ and ${\rm D}$ are connected by a BS sequence with a small number of elements. To connect each element, the corresponding radii are large, hence, the UAV trajectory is short: as shown in Fig. \ref{fig_trajectory4} (a), ${\rm S}$ and ${\rm D}$ are connected by three BSs with large radii. With the increase of the point density, $T_{\rm min}$ increases first due to some association BS sequence existing but the UAV needs to travel farther, and then decreases owing to more nearby BSs existing and connected, as shown in Fig. \ref{fig_trajectory4} (b) and (c).

\section{Conclusion}
In this paper, we consider a novel cellular network for aerial users, which is composed of dedicated BSs and traditional TBS, where the  dedicated BSs are deployed on roadside furniture. For the proposed network, we first compute the aerial coverage probability and show that the deployment of dedicated BSs improves the coverage probability in high dense urban areas and  rural areas. We then consider a cellular-connected UAV that has a flying mission and optimize its trajectory to maximize the minimum SINR. To obtain the Max-Min SINR and minimal time trajectory, we proposed two algorithms, which are more practical in large-scale networks. Finally, our results show that the optimal densities of dedicated BSs that maximize Max-Min SINR are different for different road densities. 

\appendix
\subsection{Horizontal Distances in Lemma \ref{lemma_associateprob}}\label{app_d}
The horizontal distances in Lemma \ref{lemma_associateprob} are given by:
\begin{align}
	d_{db,tb,ll}(d) &= \sqrt{\bigg(g_a(d)^{-\frac{2}{\alpha_l}}(d^2+\Delta h_{1}^2)-\Delta h_{2}^2\bigg)^{+}},\nonumber\\
	d_{db,tb,ln}(d) &= \sqrt{\bigg(\bigg(\frac{\eta_n}{\eta_{l}g_a(d)}\bigg)^{\frac{2}{\alpha_n}}(d^2+\Delta h_{1}^2)^{\frac{\alpha_{l}}{\alpha_{n}}}-\Delta h_{2}^2\bigg)^{+}}, \nonumber\\
	d_{db,db,ll}(d) &= d, \nonumber\\
	d_{db,db,ln}(d) &=\left\{ 
	\begin{aligned}
		\sqrt{\bigg(\bigg(\frac{\eta_n}{\eta_{l}}\bigg)^{\frac{2}{\alpha_n}}(d^2+\Delta h_{1}^2)^{\frac{\alpha_{l}}{\alpha_{n}}}-\Delta h_{1}^2\bigg)^{+}},  & \quad d<z_{db},\\
		\max\bigg[\sqrt{\bigg(\bigg(\frac{\eta_n}{\eta_{l}}\bigg)^{\frac{2}{\alpha_n}}(d^2+\Delta h_{1}^2)^{\frac{\alpha_{l}}{\alpha_{n}}}-\Delta h_{1}^2\bigg)^{+}},&\\\min\bigg[z_{db},\sqrt{\bigg(\bigg(\frac{g_a(d)\eta_n}{\eta_{l}}\bigg)^{\frac{2}{\alpha_n}}(d^2+\Delta h_{1}^2)^{\frac{\alpha_{l}}{\alpha_{n}}}-\Delta h_{1}^2\bigg)^{+}}\bigg]\bigg],  & \quad d>z_{db},\\
	\end{aligned} \right. 
\end{align}
\begin{align}
	d_{db,tb,nl}(d) &= \sqrt{\bigg(\bigg(\frac{\eta_l}{\eta_{n}g_a(d)}\bigg)^{\frac{2}{\alpha_l}}(d^2+\Delta h_{1}^2)^{\frac{\alpha_{n}}{\alpha_{l}}}-\Delta h_{2}^2\bigg)^{+}},\nonumber\\
	d_{db,tb,nn}(d) &= \sqrt{\bigg(g_a(d)^{-\frac{2}{\alpha_n}}(d^2+\Delta h_{1}^2)-\Delta h_{2}^2\bigg)^{+}}, \nonumber\\
	d_{db,db,nl}(d) &=\left\{ 
	\begin{aligned}
		\min\bigg[\sqrt{\bigg(\bigg(\frac{\eta_l}{\eta_{n}}\bigg)^{\frac{2}{\alpha_l}}(d^2+\Delta h_{1}^2)^{\frac{\alpha_{n}}{\alpha_{l}}}-\Delta h_{1}^2\bigg)^{+}},&\\
		\max\bigg[\sqrt{\bigg(\bigg(\frac{\eta_l}{g_a(d)\eta_{n}}\bigg)^{\frac{2}{\alpha_l}}(d^2+\Delta h_{1}^2)^{\frac{\alpha_{n}}{\alpha_{l}}}-\Delta h_{1}^2\bigg)^{+}},z_{db}\bigg]\bigg],  & \quad d<z_{db}\\
		\sqrt{\bigg(\bigg(\frac{\eta_l}{\eta_{n}}\bigg)^{\frac{2}{\alpha_l}}(d^2+\Delta h_{1}^2)^{\frac{\alpha_{n}}{\alpha_{l}}}-\Delta h_{1}^2\bigg)^{+}},  & \quad d>z_{db},\\
	\end{aligned} \right.\nonumber\\
	d_{db,db,nn}(d) &= d,
\end{align}
\begin{align}
	d_{tb,tb,ll}(d) &= d,\nonumber\\
	d_{tb,tb,ln}(d) &= \sqrt{\bigg(\bigg(\frac{\eta_n}{\eta_{l}}\bigg)^{\frac{2}{\alpha_n}}(d^2+\Delta h_{2}^2)^{\frac{\alpha_{l}}{\alpha_{n}}}-\Delta h_{2}^2\bigg)^{+}}, \nonumber\\
	d_{tb,db,ll}(d) &= \left\{ 
	\begin{aligned}
		\min\bigg[\sqrt{\bigg(g_{a}(d)^{\frac{2}{\alpha_l}}(d^2+\Delta h_{2}^2)-\Delta h_{1}^2\bigg)^{+}},z_{db}\bigg],  & \quad d<z_{db},\\
		d,  & \quad d>z_{db},\\
	\end{aligned} \right. \nonumber\\
	d_{tb,db,ln}(d) &= \min\bigg[\sqrt{\bigg(\bigg(\frac{g_a(d)\eta_n}{\eta_{l}}\bigg)^{\frac{2}{\alpha_n}}(d^2+\Delta h_{2}^2)^{\frac{\alpha_{l}}{\alpha_{n}}}-\Delta h_{1}^2\bigg)^{+}},\nonumber\\
	&\min\bigg[z_{db},\sqrt{\bigg(\bigg(\frac{\eta_n}{\eta_{l}}\bigg)^{\frac{2}{\alpha_n}}(d^2+\Delta h_{2}^2)^{\frac{\alpha_{l}}{\alpha_{n}}}-\Delta h_{1}^2\bigg)^{+}}\bigg]\bigg],
\end{align}
\begin{align}
	d_{tb,tb,nl}(d) &= \sqrt{\bigg(\bigg(\frac{\eta_l}{\eta_{n}}\bigg)^{\frac{2}{\alpha_l}}(d^2-\Delta h_{2}^2)^{\frac{\alpha_{n}}{\alpha_{l}}}-\Delta h_{1}^2\bigg)^{+}},\nonumber\\
	d_{tb,tb,nn}(d) &= d, \nonumber\\
	d_{tb,db,nl}(d) &= \min\bigg[\sqrt{\bigg(\bigg(\frac{g_a(d)\eta_l}{\eta_{n}}\bigg)^{\frac{2}{\alpha_l}}(d^2+\Delta h_{2}^2)^{\frac{\alpha_{n}}{\alpha_{l}}}-\Delta h_{1}^2\bigg)^{+}},\nonumber\\
	&\max\bigg[z_{db},\sqrt{\bigg(\bigg(\frac{\eta_l}{\eta_{n}}\bigg)^{\frac{2}{\alpha_l}}(d^2+\Delta h_{2}^2)^{\frac{\alpha_{n}}{\alpha_{l}}}-\Delta h_{1}^2\bigg)^{+}}\bigg]\bigg], \nonumber\\
	d_{tb,db,nn}(d) &= \min\bigg[\sqrt{\bigg(g_{a}(d)^{\frac{2}{\alpha_n}}(d^2+\Delta h_{2}^2)-\Delta h_{1}^2\bigg)^{+}},\max\bigg[z_{db},\sqrt{(d^2+\Delta h_{2}^2)-\Delta h_{1}^2}\bigg]\bigg],
\end{align}
where $(a)^{+}$ denotes $\max(0,a)$.

\subsection{Proof of Lemma \ref{lemma_laplace}}
\label{app_laplace}
\begin{figure}
	\centering
	\includegraphics[width=1\columnwidth]{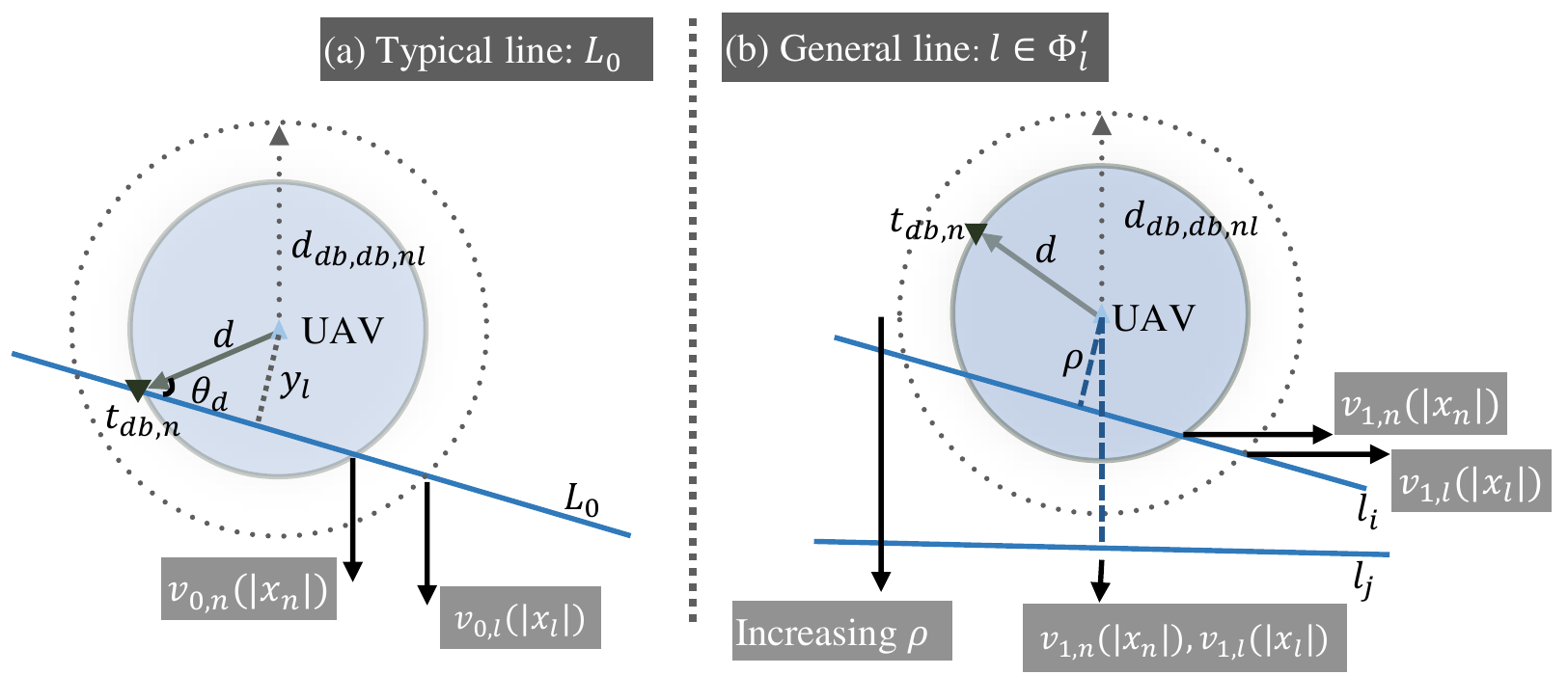}
	\caption{Illustration of the lower bound of integration in the case of associating with a dedicated BS.}
	\label{fig_I_int}
\end{figure}

The Laplace transform of the interference when associated with a dedicated BS is obtained by
\begin{align}
	\mathcal{L}_{I_{db}}(s)  =& \mathbb{E}_{I_{db}}[\exp(-s(I_{db}))] \nonumber\\
	\stackrel{(a)}{=}& \mathbb{E}_{\Phi_{db}}\bigg[\exp\bigg(-s\bigg(\sum_{db_{l,i}\in \Phi_{l}^{'}\cup L_0}p_{db}  g_a(Z_l) \eta_l G_l (Z_{l,i}^2+\Delta h_{1}^2)^{-\alpha_l/2}\nonumber\\
	&+\sum_{db_{n,j}\Phi_{l}^{'}\cup L_0}p_{db} g_a(Z_n) \eta_n G_n (Z_{n,j}^2+\Delta h_{1}^2)^{-\alpha_n/2}\bigg)\bigg)\nonumber\\
	&\times\mathbb{E}_{\Phi_{tb}}\bigg[\exp\bigg(-s\bigg(\sum_{t_{l,i}\in \Phi_{tb}}p_{tb}g_{s}\eta_l G_l D_{t_{l,i}}^{-\alpha_l}+\sum_{t_{n,j}\in \Phi_{tb}}p_{tb}g_{s}\eta_n G_n D_{t_{n,j}}^{-\alpha_n}\bigg)\bigg)\bigg]\nonumber\\
	=&\mathbb{E}_{\Phi_{db}}\bigg[\prod_{db_{l,i}\in \Phi_{l}^{'}\cup L_0}\exp\bigg(-s p_{db} g_a(Z_l) \eta_l G_l (Z_{l}^2+\Delta h_{1}^2)^{-\alpha_l/2}\bigg)\nonumber\\
	&\times
	\prod_{db_{n,j}\Phi_{l}^{'}\cup L_0} \exp\bigg(-s p_{db}g_{s} g_a(Z_n) \eta_n G_n (Z_{n}^2+\Delta h_{1}^2)^{-\alpha_n/2}\bigg)\bigg]
	\nonumber\\
	&\times
	\mathbb{E}_{\Phi_{tb}}\bigg[\prod_{t_{l,i}\in \Phi_{tb}}g_{s}\exp\bigg(-s p_{tb}g_{s}\eta_l G_l D_{t_l}^{-\alpha_l}\bigg)
	\prod_{t_{n,j}\in \Phi_{tb}}\exp\bigg(-s p_{tb}g_{s}\eta_n G_n D_{t_n}^{-\alpha_n}\bigg)\bigg]\nonumber\\
	=& \mathbb{E}_{\Phi_{db}}\bigg[\prod_{db_{l,i}\Phi_{l}^{'}\cup L_0}\bigg(\frac{m_l}{m_l+s p_{db} g_a(Z_l) \eta_l (Z_{l}^2+\Delta h_{1}^2)^{-\alpha_l/2}}\bigg)^{m_l}	\nonumber\\
	&\times
	\prod_{db_{n,j}\Phi_{l}^{'}\cup L_0}\bigg(\frac{m_n}{m_n+s  p_{db} g_a(Z_n) \eta_n G_n (Z_{n}^2+\Delta h_{1}^2)^{-\alpha_n/2}}\bigg)^{m_n}\bigg]
	\nonumber\\
	&
	\times\mathbb{E}_{\Phi_{tb}}\bigg[\prod_{t_{l,i}\in \Phi_{tb}}\bigg(\frac{m_l}{m_l+s p_{tb}g_{s}\eta_l D_{t_l}^{-\alpha_l}}\bigg)^{m_l} 
	\prod_{t_{n,j}\in \Phi_{tb}}\bigg(\frac{m_n}{m_n+s p_{tb}g_{s}\eta_n D_{t_n}^{-\alpha_n}}\bigg)^{m_n}\bigg],
\end{align}
where step (a) follows from we write the typical line with general lines together, and proof completes by using the probability generating functional of inhomogeneous PPP and PLCP. Just notice that for the interference from general lines, we need to compute the lines intersect with $B({\rm UAV,d_{db,db,nn}})$, ($\rho<d_{db,db,nn}$), and lines do not intersect with $B({\rm UAV,d_{db,db,nl}})$, ($\rho>d_{db,db,nn}$), separately for the NLoS interference, and the lines intersect with $B({\rm UAV,d_{db,db,nn}})$, ($\rho<d_{db,db,nn}$), and lines do not intersect with $B({\rm UAV,d_{db,db,nl}})$, ($\rho>d_{db,db,nn}$), separately for the LoS interference,  where $B(a,b)$ denotes the ball centered at $a$ with radius $b$, as shown in Fig. \ref{fig_I_int}.

The Laplace transform $\mathcal{L}_{I_{tb}}$ follows a similar method and omits the
the interference from the typical line, thus, omitted here.

%\subsection{Proof of Lemma \ref{lemma_successprob}}
%\label{app_successprob}
%The success probability in the case of associating with a LoS dedicated BS is given by
%\begin{align}
%	P_{s,td_l}(\tau,d) &= \mathbb{P}\bigg(\frac{p_t \eta_l g_{a}(d) G_l (d^2+\Delta h_{1}^2)^{-\frac{\alpha_l}{2}}}{I+\sigma^2}>\tau\bigg)\nonumber\\
%	&= \mathbb{P}\bigg(G_l>\tau(I+\sigma^2)p_t \eta_l g_{a}(d) (d^2+\Delta h_{1}^2)^{-\frac{\alpha_l}{2}}\bigg) \nonumber\\
%	&\stackrel{(a)}{\approx}\sum_{k = 1}^{m_l}\binom{k}{m_l}(-1)^{k+1}\exp(-k \beta_2(m_l)m_l \tau (I+\sigma^2)(p_t \eta_l g_{a}(d))^{-1} (d^2+\Delta h_{1}^2)^{\frac{\alpha_l}{2}})\nonumber\\
%	&= \sum_{k = 1}^{m_l}\binom{k}{m_l}(-1)^{k+1}\mathcal{L}_{I+\sigma^2}(k \beta_2(m_l)m_l \tau (p_t \eta_l g_{a}(d))^{-1} (d^2+\Delta h_{1}^2)^{\frac{\alpha_l}{2}}),
%\end{align}
%where step (a) follows from the upper bound approximation of Gamma distribution and can be found in \cite{}.
	\bibliographystyle{IEEEtran}
	\bibliography{Rep13}
\end{document}